\pdfoutput=1 

\documentclass[twocolumn]{aastex62}

\usepackage{hyperref}
\usepackage{amsmath,amstext}
\usepackage[T1]{fontenc}
\usepackage{float}

\usepackage{natbib}
\usepackage[figure,figure*]{hypcap}
\usepackage{graphicx}
\usepackage{tablefootnote}
\usepackage{multirow}
\usepackage{tabularx}

\usepackage{array}
\usepackage{microtype}

\shortauthors{}

\begin{document}

\title{No sign of G2's encounter affecting Sgr A*'s X-ray flaring rate from $Chandra$ observations}

\correspondingauthor{\'Elie Bouffard}
\email{elie.bouffard@mail.mcgill.ca}

\author{\'Elie Bouffard}
\author{Daryl Haggard}
\affil{McGill Space Institute and Department of Physics, McGill University, 3600 rue University, Montreal, QC, H3A 2T8, Canada}
\author{Michael A. Nowak}
\affil{Department of Physics, Washington University, One Brookings Drive, St. Louis, MO 63130-4899, USA}
\author{Joseph Neilsen}
\affil{Villanova University, Mendel Hall, Room 263A, 800 E. Lancaster Avenue, Villanova, PA 19085, USA}
\author{Sera Markoff}
\affil{Anton Pannekoek Institute for Astronomy, University of Amsterdam, Science Park 904, 1098 XH, Amsterdam, The Netherlands}

\author{Frederick K. Baganoff}
\affil{MIT Kavli Institute for Astrophysics and Space Research, Cambridge, MA 02139, USA}

\begin{abstract}
An unusual object, G2, had its pericenter passage around Sgr A*, the $4\times10^6$ M$_\odot$ supermassive black hole in the Galactic Centre, in Summer 2014. Several research teams have reported evidence that following G2's pericenter encounter the rate of Sgr A*'s bright X-ray flares increased significantly. Our analysis carefully treats varying flux contamination from a nearby magnetic neutron star and is free from complications induced by using data from multiple X-ray observatories with different spatial resolutions. We test the scenario of an increased bright X-ray flaring rate using a massive dataset from the \textit{Chandra X-ray Observatory}, the only X-ray instrument that can spatially distinguish between Sgr A* and the nearby Galactic Centre magnetar throughout the full extended period encompassing G2's encounter with Sgr A*. We use X-ray data from the 3 Ms observations of the \textit{Chandra} \textit{X-ray Visionary Program} (XVP) in 2012 as well as an additional 1.5 Ms of observations up to 2018. We use detected flares to make distributions of flare properties. Using simulations of X-ray flares accounting for important factors such as the different $Chandra$ instrument modes, we test the null hypothesis on Sgr A*'s bright (or any flare category) X-ray flaring rate around different potential change points. In contrast to previous studies, our results are consistent with the null hypothesis; the same model parameters produce distributions consistent with the observed ones around any plausible change point.

\end{abstract}

\section{Introduction}
\label{sec:intro}
At a distance of $\sim$ 8 kpc, in the centre of our Milky Way galaxy, lies a $\sim$ $4\times 10^{6} \text{ M}_{\odot}$ \citep{gillessen2017update,abuter2018detection} supermassive black hole, Sgr A*. It is radiating very faintly at a bolometric luminosity about 9  orders of magnitude lower than its Eddington luminosity \citep{genzel2010galactic}. In the 2-10 keV X-ray band, Sgr A*'s behaviour is characterized by a constant, quiescent unabsorbed luminosity of $\sim$ 3.4 $\times$ $10^{33}$ erg s$^{\text{-1}}$\citep{wang2013dissecting}. This emission is spatially extended and well modeled by thermal bremsstrahlung emission arising from thermal plasma in the accretion flow around the Bondi radius \citep{quataert2002thermal,baganoff2003chandra,yuan2003nonthermal,liu2004electron,xu2006thermal,wang2013dissecting}. This quiescent behaviour is punctuated by X-ray flares of luminosities varying from $\sim$ 10s to 100s of times quiescence \citep[or 100s to 1000s of times their locally-emitted background since the inner accretion flow contributes 10\% of the quiescent emission;][]{baganoff2001rapid,goldwurm2003new,porquet2003xmm,porquet2008x,belanger2005repeated,nowak2012chandra,neilsen2013chandra,neilsen2015x,degenaar2013x,barriere2014nustar,ponti2015fifteen,mossoux2016multiwavelength,yuan2015systematic,zhang2017sagittarius,yuan2017systematic, boyce2019simultaneous,Haggard19}. 

Several studies have been performed to understand the statistical behaviour of Sgr A*'s X-ray flares. The first by \citet{neilsen2013chandra} was done before G2 and used the 3 Ms of \textit{Chandra} data from the 2012 \textit{X-ray Visionary Project} (XVP\footnote{The Co-PIs are Frederick K. Baganoff, Sera Markoff and Michael A. Nowak.}). They found 39 flares with $L_{2-8 keV} > 10^{34}$ erg s$^{\text{-1}}$ corresponding to a constant flaring rate of $1.1^{+0.2}_{-0.1}$ flares per day using a Gaussian fitting technique on the binned 300 s 2-8 keV light curves. 

\citet{yuan2015systematic} and \citet{yuan2017systematic} also performed an extensive X-ray flare study, this time on 4.5 Ms of \textit{Chandra} observations from 1999 to 2012. They detected 82 flares, including 49 in the XVP dataset, using a maximum likelihood fitting algorithm on the unbinned 2-8 keV light curves. Their results are mostly consistent with \citet{neilsen2013chandra}.

Possible sources of flares include magnetic reconnection, stochastic acceleration, shocks from jets or the accretion flow, or even tidal disruption of asteroids \citep{markoff2001nature,liu2002accretion,yuan2003nonthermal,liu2004electron,vcadevz2008tidal,kostic2009tidal,zubovas2012sgr,dibi2014exploring,dibi2016using,ball2016particle,ball2018properties} whereas the responsible radiation mechanisms are likely synchrotron or synchrotron self-Compton in nature \citep{marrone2008x,dodds2009evidence,eckart2009modeling,witzel2012source,nowak2012chandra,barriere2014nustar,neilsen2015x,ponti2017powerful,zhang2017sagittarius} but not inverse-Compton where the photons that get up-scattered to X-rays come from an external region \citep{boyce2019simultaneous}. Flares share a consistent spectrum with a photon index $\Gamma \sim 2$ \citep{nowak2012chandra,neilsen2013chandra,ponti2017powerful,yuan2017systematic} and their timescale of minutes to hours points to an origin at $\sim 10$'s of Schwarzschild radii \citep[R$_s$;][]{quataert2003radiatively}. This scale was recently confirmed by IR observations from the GRAVITY Collaboration, which tracked 3 flares as they orbited around Sgr A* \citep{abuter2018detection}.

The work of \citet{boyce2019simultaneous} shows that X-ray flares lead the IR variations by 10-20 minutes but is also consistent with no time lag at all. These authors cannot pick a single model that matches all observations. Luckily, there will be joint $Spitzer/Chandra/$GRAVITY observations in Summer 2019 which could help break this degeneracy. These exceptional multiwavelength observations will push our understanding of Sgr A*'s accretion physics further than ever before.  

Furthermore, data from the Event Horizon Telescope (EHT) Campaigns capable of resolving Sgr A* and M87's black holes \citep[e.g.,][]{EHTm87-I} will put strong constraints on the underlying physical processes behind flares and the accretion flow, and inform better general relativistic magnetohydrodynamic simulations.

In 2014, G2, a mysterious object of $\sim$ 3 Earth masses on a very eccentric orbit of $e \sim 0.97$ reached a pericentre of $\sim 2000 \text{ R}_s$ around Sgr A* \citep{gillessen2012gas,gillessen2013new, gillessen2013pericenter}, probing a region of the accretion flow previously unconstrained by observations. Different research groups postdict different pericentre passage times for G2 depending on their proposed models. For example, \citet{madigan2016using} show that a pure Keplerian orbit would have given an estimated peripassage time between the end of February and mid April 2014 but that the addition of a drag force would have pushed this date to between the end of May and mid-July of the same year. If an inflow was also added from the accretion flow, the pericenter passage time would have become between early July and early September 2014. \citet{gillessen2018detection} propose a model involving a drag force without an inflow. They compare their orbital fit to one using only Keplerian motion and find that the drag force was significant at the 10$\sigma$ level. They report a pericentre passage time of 2014.58 $\pm$ 0.13, earlier but consistent with \citet{madigan2016using}. 

This encounter could have resulted in an enhanced rate and luminosity of Sgr A*'s flares due to an increased accretion rate. Bright flares could have been generated if G2's material was clumpy, leading to accretion in bunches, or by shocks following the interaction between the cloud and the accretion flow \citep{ponti2015fifteen}.

\citet{schartmann2012simulations} and \citet{kawashima2017possible} predict an increase in activity between a few years and 5-10 years after pericentre, respectively. Whereas the former explain this delay by the strong angular momentum of G2 (which delays accretion), the latter use full 3D general relativistic magnetohydrodynamic simulations to model the evolution of magnetic fields as they interact with the cloud. An instability develops, and a magnetic reconnection event is expected to increase the radio and X-ray luminosity on a dynamo-viscous timescale of 5-10 years.

Sgr A* was closely monitored in Summer and Fall 2014 to detect any change in activity related to G2. Using $Chandra$, \textit{XMM}\textit{-}\textit{Newton} and $Swift$ observations, multiple groups each report an increase in the bright flaring rate of Sgr A*.

\citet{ponti2015fifteen} report 80 flares from $Chandra$ and \textit{XMM}\textit{-}\textit{Newton} in 6.9 Ms of data from September 1999 to November 2014. They find an increase in the bright/very bright flaring rate (defined as flares with an absorbed fluence greater than $5 \times 10^{-9}$ erg cm$^{-2}$) from $0.27 \pm 0.04$ to $2.5 \pm 1.0$ per day at 99.9\% confidence after summer 2014. They also report a decline in the moderate-bright flaring rate from mid-2013 at the 96\% level. 

\citet{mossoux2017sixteen} use 9.3 Ms of data from 1999 to 2015 from $Chandra$, \textit{XMM}\textit{-}\textit{Newton} and $Swift$, and report 107 flares. They also find an increase in the most energetic flaring rate by a factor of 3 following 2014 August 31 and a decay for the faintest flares by a similar factor. To do so, they use bottom-to-top and top-to-bottom searches on the detected flare fluxes and fluences. The bottom-to-top search consists on removing the faintest flare then performing the search, and repeating until a signal is found. The top-to-bottom search follows the same logic but the brightest flare is removed at each step instead. 

Both of these works use data from $Chandra$ but also from \textit{XMM}\textit{-}\textit{Newton} and $Swift$ to have as many flares as possible. This complicates the analysis since all these observatories have different sensitivities and spatial resolutions. These authors also do not perform extensive Monte Carlo simulations of X-ray flares. \citet{mossoux2017sixteen} do simulate X-ray flares to control their detection bias and they also allow the simulated flares to start before the beginning or after the end of a given observation to consider edge effects. However, they do not use their tool to study the impact of other factors such as multiple overlapping flares..

In this work, we use our $Chandra$ observations of the Galactic Centre to re-visit the question of whether there was an increase in Sgr A*'s X-ray flare rate near G2's pericenter passage. We use Bayesian Blocks \citep{scargle2013studies,williams2017pwkit} to detect and characterize flares, and we create emitted energy and duration distributions of flares similarly to \citet{neilsen2013chandra}. By splitting the observations into two datasets around a potential change point related to G2 (and repeating this process for every potential change point), we test the null hypothesis by performing Monte Carlo simulations of each X-ray light curve within each dataset using a model with the same parameters to produce confidence intervals for the flare distributions of both datasets. 

This paper is organized as follows. In Section \ref{sec:obs} we explain the $Chandra$ data reduction. In Section \ref{sec:BB} we explain how we use Bayesian Blocks to detect and characterize flares, and we show our detected flares from the XVP and Post-XVP datasets. In Section \ref{sec:simulation}, we present our Monte Carlo simulation model and look for significant change points in the X-ray flare rate. We discuss our results in Section \ref{sec:Discus} and conclude in Section \ref{sec:Conclu}.

\section{Observations}
\label{sec:obs}
\subsection{Chandra X-ray Visionary Program}
$Chandra$ collected 38 observations of Sgr A* during the 2012 XVP campaign\footnote{\url{https://www.sgra-star.com/}}, for a total of $\sim$3 Ms of data. The observations were all taken with the High Energy Transmission Grating (HETG) Spectrometer with the Advanced CCD Imaging Spectrometer-Spectroscopy (ACIS-S) camera at focus. This instrument mode has a frame time of 3.14 s and the HETG disperses some of the photons across the detectors to increase spectral resolution.

We reduce these data using \texttt{CIAO v.4.9} tools (\texttt{CALDB v.4.7.6}) and we reprocess the level 2 events with the \texttt{chandra\_repro} script before updating the WCS coordinate system with the \texttt{wcs\_update} tool. The diffraction order of each event is determined by the \texttt{tg\_resolve\_events} tool and we keep zeroth and $\pm$first order photons. We use the same extraction region as \citet{nowak2012chandra}, \citet{neilsen2013chandra} and \citet{mossoux2017sixteen} to minimise background. We extract 2-8 keV zeroth-order events from a small 1.25" radius circular region centered around the radio position of the Sgr A* \citep[17$^{\text{h}}$45$^{\text{m}}$40$^{\text{s}}$.04909, -29$^{\circ}$00$^{\text{'}}$28.118$^{\text{"}}$;][]{reid2004proper} and first-order counts (with grating order tolerance of $\pm 0.2$) are extracted from a rectangular box with a 5 pixel (2.5") width also centered on the source. The PHA2 files and the gratings responses are created with the \texttt{tg\_extract} and \texttt{mktgresp} tools, respectively, while the zeroth order spectra and response files are extracted with the \texttt{specextract} tool.

\subsection{Post-XVP data}
We select $Chandra$ observations following the 2012 XVP campaign from 2013 May 12 to 2018 April 25. To create a consistent dataset, we only use observations from the ACIS-S3 1/8th subarray instrument mode. This instrument mode reduces pile-up by using a 1/8th subarray of the central S3 chip (128-rows), which reduces the frame time to 0.44 s. 

We exclude ObsIDs 14944 and 16597 because \texttt{wcs\_match} did not properly correct the WCS coordinate system due to their short exposure (20 ks and 18 ks) making the localization of the sources more difficult. Inaccurate positions of the extraction regions would significantly affect the contribution of the magnetar to Sgr A*'s quiescent flux (see Section \ref{sec:magn}). Similarly, we exclude ObsIDs 18055 and 18056 because of the presence of a low-mass X-ray binary in outburst, Transient 15 \citep{ponti2016swift}. Our final Post-XVP ACIS-S3 1/8th subarray dataset consists of 1.56 Ms of data across 39 observations.

\begin{figure}[h!]
    \centering
    \includegraphics[width=0.47\textwidth]{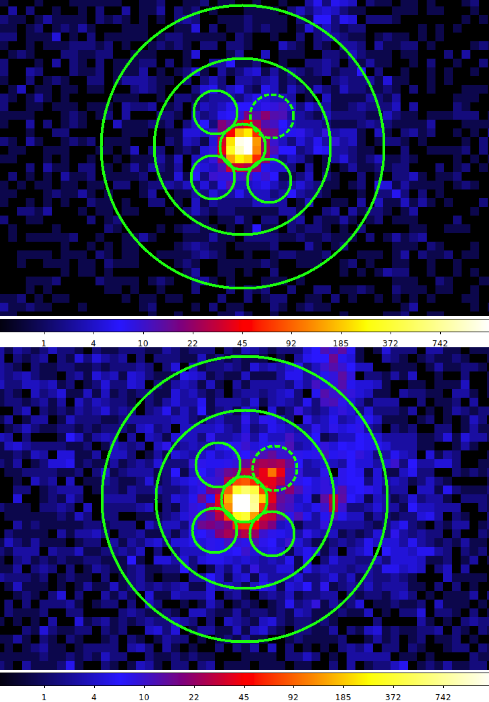} 
    \caption{$Top$: Image of ObsID 14703 (2013 June 4) from $\textit{Chandra}$. The annulus is the background extraction region used for each source (inner radius of 5" and outer radius of 8", center on RA:17$^{\text{h}}$45$^{\text{m}}$40$^{\text{s}}$.084, DEC:-29$^{\circ}$00$^{\text{'}}$28.70$^{\text{"}}$). The brightest source is the magnetar (extraction region centered on RA:17$^{\text{h}}$45$^{\text{m}}$40$^{\text{s}}$.169, DEC:-29$^{\circ}$00$^{\text{'}}$29.84$^{\text{"}}$ with a radius of 1.3"). The dashed circle towards the upper-right from the magnetar is the region for Sgr A* (extraction region centered on RA:17$^{\text{h}}$45$^{\text{m}}$40$^{\text{s}}$.0409, DEC:-29$^{\circ}$00$^{\text{'}}$28.118$^{\text{"}}$ with a radius of 1.25"). The three other regions have the same radius as Sgr A* and their center's position are : RA:17$^{\text{h}}$45$^{\text{m}}$40$^{\text{s}}$.2971, DEC:-29$^{\circ}$00$^{\text{'}}$31.57$^{\text{"}}$ and RA:17$^{\text{h}}$45$^{\text{m}}$40$^{\text{s}}$.0542, DEC:-29$^{\circ}$00$^{\text{'}}$31.7615$^{\text{"}}$ and RA:17$^{\text{h}}$45$^{\text{m}}$40$^{\text{s}}$.2838 and DEC:-29$^{\circ}$00$^{\text{'}}$27.9185$^{\text{"}}$. $Bottom$: Same as before but for ObsID 15042 (2013 August 11) during which Sgr A* flared.}
    \label{fig:ObsID14703}
\end{figure}

We adopt a 1.25" radius circular extraction region around Sgr A* to select 2-8 keV events after reprocessing the data with the \texttt{chandra\_repro} tool and updating the WCS coordinate system with \texttt{wcs\_update}. We extract X-ray spectra with the \texttt{specextract} tool. There is one complication, however; on 2013 April 25, SGR J1745-29, a new magnetar located only 2.4" from Sgr A*, went into outburst \citep[see Figure \ref{fig:ObsID14703};][]{mori2013nustar,kennea2013swift}. Its luminosity is so great that its PSF leaks into Sgr A*'s extraction region. This influences how we handle background for the majority of this dataset (see Section \ref{sec:magn}). 

\subsection{Magnetar contamination}
\label{sec:magn}
The top of Figure \ref{fig:ObsID14703} shows a $\textit{Chandra}$ image taken on 2013 June 4 in which the magnetar is much brighter than Sgr A*. The bottom of the same figure shows another observation taken on 2013 August 11 during which Sgr A* flared. The magnetar only drops to a luminosity comparable (within a factor of 2) to Sgr A*'s quiescence about 3 years after outburst. This means that this bright source contaminates our Sgr A* extraction region for the majority of our Post-XVP data. There is no need to apply this special data treatment to data taken before the outburst since the magnetar is too faint to cause any sort of contamination. 

The analysis presented in this Section is a key element for simulating quiescent count rates in Section \ref{sec:simulation}.

\subsubsection{Estimation of the magnetar's contamination}
\label{subsec:magcont}

To isolate Sgr A*'s quiescence count rate, we must calculate the subset of the magnetar's counts that overlap Sgr A*'s extraction region. We adopt the background region from \citet{2017MNRAS.471.1819C}, i.e., an annulus with an inner radius of 5" and outer radius of 8" (see Figure \ref{fig:ObsID14703}). We extract count rates in the 2-8 keV band from 3 regions located at the same radial distance from the magnetar as Sgr A*, with the same size as Sgr A*'s extraction region (radius of 1.25"). We take the mean of the extracted, background-subtracted count rates and divide the result by the 2-8 keV count rate of the magnetar's extraction region (radius of 1.3"). This results in the fraction $\epsilon$ of the count rate $Q_{\text{magn}}$ from the magnetar that leaks out to Sgr A*'s radial distance. The measured Sgr A* quiescence count rate $Q_{\text{eff}}$ (corresponding to the longest block obtained with Bayesian Blocks, see Section \ref{sec:BB_class}) is in fact the sum of the \textit{actual} Sgr A* quiescence count rate $Q_{\text{sgr}}$ and a magnetar contribution $\epsilon \times Q_{\text{mag}}$  

\begin{equation}
\label{eq:magn_frac}
    Q_{\text{eff}} = \epsilon \times Q_{\text{mag}} + Q_{\text{sgr}}
\end{equation}
\noindent
We compute $\epsilon$ for each of our Post-XVP observations until July 12 2016 (ObsID 18731) since at that point the magnetar is faint enough that count rates in the testing regions fall to background level.

\begin{figure}[h!]
    \centering
    \includegraphics[width=0.47\textwidth]{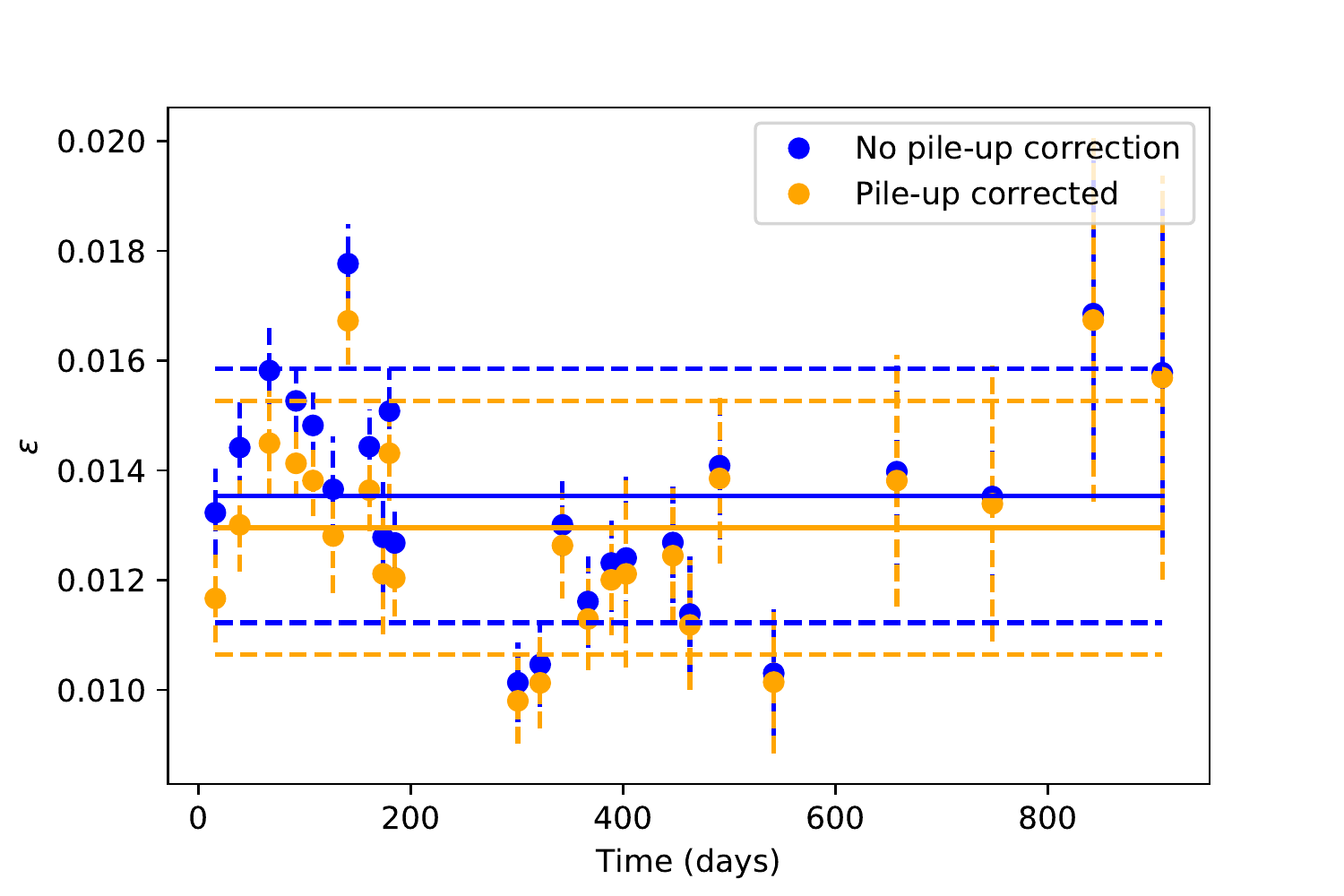}
    \caption{Each point corresponds to an ObsID. Error bars are statistical Poisson errors obtained from the counts. The horizontal lines are the mean value of their respective dataset (pile-up corrected in orange and not, in blue) obtained via Monte Carlo simulations. Their associated $\pm 1\sigma$ region is delimited by horizontal dashed lines of the same colour. We obtain $\left \langle \epsilon \right \rangle= (1.4 \pm 0.2) \%$ without treating pile-up and $\left \langle \epsilon \right \rangle= (1.3 \pm 0.2) \%$ if we treat it.}
    \label{fig:Magn_frac}
\end{figure}

The resulting plot of $\epsilon$ as a function of time since the outburst is shown in Figure \ref{fig:Magn_frac}. The observed fluctuation of $\epsilon$ can be explained with possible deviations from observation to observation in source localization combined with Poisson noise on the counts. We show the statistical Poisson error bars on each value of $\epsilon$. Since we do not observe any trend over long periods, we conclude that those variations are random and proceed to calculate the global mean and standard deviation of $\epsilon$ using Monte Carlo simulations. For each ObsID, we draw a random point from its associated value and standard deviation. Once we have done this for every point, we save the mean and the standard deviation of $\epsilon$. We repeat this process 10 000 times and take the average of all the saved means and standard deviations. We obtain $\left \langle \epsilon \right \rangle = (1.4 \pm 0.2) \%$ without correcting for pile-up and $\left \langle \epsilon \right \rangle = (1.3 \pm 0.2)\%$ if we correct for it. Pile-up decreases $\left \langle \epsilon \right \rangle$ because it only affects the magnetar given the low rates of the other regions. The following analysis will use the $\epsilon$ obtained with a pile-up correction. 

\subsubsection{Implication for Sgr A*'s quiescent count rate}
With $\epsilon$, we can estimate Sgr A*'s quiescent count rate $Q_{\text{sgr}}$ using Equation \ref{eq:magn_frac} for every observation. The associated error on each $Q_{\text{sgr}}$ is found by propagating the uncertainties of the parameters in Equation \ref{eq:magn_frac}.

To find the mean quiescent count rate from this distribution, we also use Monte Carlo simulations. We obtain $\left \langle Q_{\text{sgr}} \right \rangle  = (0.005 \pm 0.001)$ ct s$^{\text{-1}}$ for Sgr A*'s average quiescent count rate.

\begin{figure}[h!]
    \centering
    \includegraphics[width=0.47\textwidth]{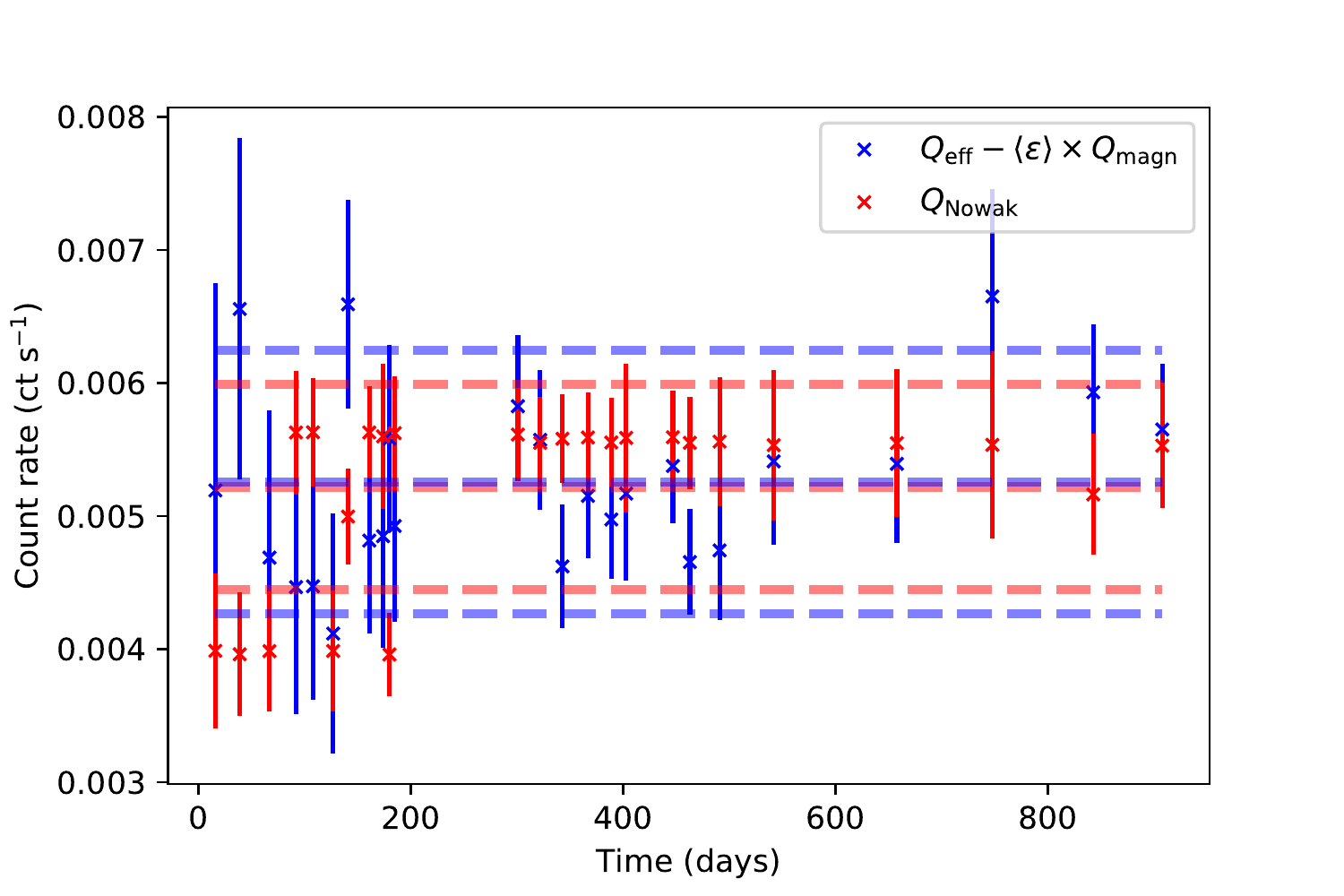}
    \caption{We plot the difference between $Q_{\text{eff}}$ and $\left \langle \epsilon \right \rangle \times Q_{\text{magn}}$ in blue and the quiescent count rate predicted by \citet{nowak2012chandra} in red, $Q_{\text{Nowak}}$, at each ObsID present in Figure \ref{fig:Magn_frac}. The error bar of each blue point is found by propagating the errors of the parameters of Equation \ref{eq:magn_frac}. The mean count rate $\left \langle Q_{\text{sgr}} \right \rangle = (0.005 \pm 0.001)$ ct s$^{\text{-1}}$ is represented by the middle blue dashed horizontal line, with its associated $\pm 1\sigma$ Monte Carlo region delimited by the other two horizontal dashed lines in the same color. The middle red dashed line shows $\left \langle Q_{\text{Nowak}}\right \rangle = (0.0052 \pm 0.008)$ ct s$^{\text{-1}}$. Its corresponding $\pm 1\sigma$ Monte Carlo region is delimited by the other two horizontal dashed lines in the same color, and the error bar of each red point is the Poisson error. }
    \label{fig:Quiescences}
\end{figure}

We compare our results with the quiescence count rates predicted using the model presented in \citet{nowak2012chandra}. Using this model in \texttt{XSPEC} \citep{arnaud1996xspec}, we compute the predicted count rates $Q_{\text{Nowak}}$ for each ObsID using the instrument response files and compare them to the expected value, $Q_{\text{sgr}}$. A comparison is shown in Figure \ref{fig:Quiescences} and Table \ref{tab:Magn}. Our resulting mean quiescent count rate $\left \langle Q_{\text{sgr}} \right \rangle$ is consistent with \citet{nowak2012chandra}'s prediction of $\left \langle Q_{\text{Nowak}} \right \rangle=(0.0052 \pm 0.0008)$ ct s$^{\text{-1}}$. Small variations in $Q_{\text{Nowak}}$ are caused by changes in instrument response files. In the following work, we choose to use our own quiescent value.

\begin{table*}[htp]
\renewcommand{\arraystretch}{0.9}
\centering
\begin{tabular}{| c | c |c|c|c|c|c|}
\hline
ObsID     & Date & Time  & Magnetar  & Detected & Predicted & $Q_{\text{sgr}}$ = \\
& & since & contribution & quiescent cr& quiescent cr & $Q_{\text{eff}}$ -  \\
& & outburst & $\left \langle \epsilon \right \rangle \times Q_{\text{mag}}$ & $Q_{\text{eff}}$ & $Q_{\text{Nowak}}$ & $\left \langle \epsilon \right \rangle \times Q_{\text{mag}}$ \\ \hline
--      &--&Days   &  $\times 10^{-3}$ ct s$^{\text{-1}}$   &   $\times 10^{-3}$ ct s$^{\text{-1}}$         &  $\times 10^{-3}$ ct s$^{\text{-1}}$& $\times 10^{-3}$ ct s$^{\text{-1}}$  \\ \hline
$14702$ & 2013 May 12 & $16  $& $7   \pm 1     $& $12  \pm 1  $ & $4.0 \pm 0.6$&$ 5 \pm 2$\\
$14703$ & 2013 Jun 04 & $39  $& $6   \pm 1     $& $12.1\pm 0.8$ & $4.0 \pm 0.5$&$ 7 \pm 2$ \\
$14946$ & 2013 Jul 02 & $67  $& $4.8 \pm 0.9   $& $9.5 \pm 0.7$ & $4.0 \pm 0.5$&$ 5 \pm 1$\\
$15041$ & 2013 Jul 27 & $92  $& $4.3 \pm 0.8   $& $8.7 \pm 0.6$ & $5.6 \pm 0.5$&$ 4 \pm 1$ \\
$15042$ & 2013 Aug 11 & $108 $& $3.9 \pm 0.7   $& $8.4 \pm 0.5$ & $5.6 \pm 0.4$&$ 4.5 \pm 0.9$\\
$14945$ & 2013 Aug 31 & $127 $& $3.6 \pm 0.6   $& $7.7 \pm 0.6$ & $4.0 \pm 0.5$&$ 4.1 \pm 0.9$ \\
$15043^{\text{a}}$ & 2013 Sep 14 & $141 $& $3.3 \pm 0.6   $& $9.9 \pm 0.5$ & $5.0 \pm 0.4$&$ 6.6 \pm 0.8$\\
$15044$ & 2013 Oct 04 & $161 $& $3.1 \pm 0.6   $& $8.0 \pm 0.4$ & $5.6 \pm 0.3$&$ 4.8 \pm 0.7$ \\
$14943$ & 2013 Oct 17 & $174 $& $3.0 \pm 0.5   $& $7.9 \pm 0.6$ & $5.6 \pm 0.5$&$ 4.9 \pm 0.8$\\
$14704^{\text{b}}$ & 2013 Oct 23 & $180 $& $3.0 \pm 0.5   $& $8.5 \pm 0.5$ & $4.0 \pm 0.3$&$ 5.6 \pm 0.7$ \\
$15045$ & 2013 Oct 28 & $185 $& $2.9 \pm 0.5   $& $7.8 \pm 0.5$ & $5.6 \pm 0.4$&$ 4.9 \pm 0.7$\\
$16508$ & 2014 Feb 21 & $301 $& $1.9 \pm 0.3   $& $7.7 \pm 0.4$ & $5.6 \pm 0.4$&$ 5.8 \pm 0.5$ \\
$16211$ & 2014 Mar 14 & $322 $& $1.9 \pm 0.3   $& $7.4 \pm 0.4$ & $5.5 \pm 0.3$&$ 5.6 \pm 0.5$\\
$16212$ & 2014 Apr 04 & $343 $& $1.7 \pm 0.3   $& $6.3 \pm 0.4$ & $5.6 \pm 0.3$&$ 4.6 \pm 0.5$ \\
$16213$ & 2014 Apr 28 & $367 $& $1.6 \pm 0.3   $& $6.7 \pm 0.4$ & $5.6 \pm 0.3$&$ 5.2 \pm 0.5$\\
$16214$ & 2014 May 20 & $389 $& $1.5 \pm 0.3   $& $6.4 \pm 0.4$ & $5.6 \pm 0.3$&$ 5.0 \pm 0.4$ \\
$16210$ & 2014 Jun 03 & $403 $& $1.4 \pm 0.2   $& $6.2 \pm 0.6$ & $5.6 \pm 0.5$&$ 5.2 \pm 0.7$\\
$16215$ & 2014 Jul 16 & $447 $& $1.1 \pm 0.2   $& $6.5 \pm 0.4$ & $5.6 \pm 0.3$&$ 5.4 \pm 0.4$ \\
$16216$ & 2014 Aug 02 & $463 $& $1.0 \pm 0.2   $& $5.7 \pm 0.4$ & $5.5 \pm 0.3$&$ 4.7 \pm 0.4$\\
$16217$ & 2014 Aug 30 & $491 $& $1.0 \pm 0.2   $& $5.7 \pm 0.5$ & $5.6 \pm 0.5$&$ 4.7 \pm 0.5$ \\
$16218$ & 2014 Oct 20 & $542 $& $0.9 \pm 0.2   $& $6.3 \pm 0.6$ & $5.5 \pm 0.6$&$ 5.4 \pm 0.6$\\
$16963$ & 2015 Feb 13 & $658 $& $0.7 \pm 0.1   $& $6.1 \pm 0.6$ & $5.5 \pm 0.6$&$ 5.4 \pm 0.6$ \\
$16966$ & 2015 May 14 & $748 $& $0.5 \pm 0.1   $& $7.2 \pm 0.8$ & $5.5 \pm 0.7$&$ 6.7 \pm 0.8$\\
$16965$ & 2015 Aug 17 & $843 $& $0.42\pm 0.08  $& $6.3 \pm 0.5$ & $5.2 \pm 0.5$&$ 5.9 \pm 0.5$ \\
$16964$ & 2015 Oct 21 & $908 $& $0.31\pm 0.06  $& $6.0 \pm 0.5$ & $5.5 \pm 0.5$&$ 5.7 \pm 0.5$\\
$18731$ & 2016 Jul 12 & $1173$& $0.15\pm 0.03  $& $5.5 \pm 0.3$ & $5.5 \pm 0.3$& -- \\
$18732$ & 2016 Jul 18 & $1179$& $0.17\pm 0.03  $& $5.0 \pm 0.3$ & $5.5 \pm 0.3$& --\\
$18057$ & 2016 Oct 08 & $1261$& $0.15\pm 0.03  $& $5.2 \pm 0.5$ & $5.5 \pm 0.5$& -- \\ 
$18058$ & 2016 Oct 14 & $1267$& $0.15\pm 0.03  $& $4.6 \pm 0.4$ & $5.6 \pm 0.5$&  -- \\
$19726$ & 2017 Apr 06 & $1441$& $0.10\pm 0.02  $& $5.0 \pm 0.4$ & $5.5 \pm 0.4$&  -- \\
$19727$ & 2017 Apr 07 & $1442$& $0.10\pm 0.02  $& $5.7 \pm 0.5$ & $5.5 \pm 0.5$&  -- \\
$20041$ & 2017 Apr 11 & $1446$& $0.12\pm 0.02  $& $7.2 \pm 0.7$ & $5.5 \pm 0.6$&  -- \\
$20040$ & 2017 Apr 12 & $1447$& $0.10\pm 0.02  $& $5.5 \pm 0.4$ & $5.5 \pm 0.4$&  -- \\
$19703$ & 2017 Jul 15 & $1541$& $0.09\pm 0.02  $& $4.6 \pm 0.3$ & $5.5 \pm 0.3$&  -- \\
$19704$ & 2017 Jul 25 & $1551$& $0.09\pm 0.02  $& $5.0 \pm 0.2$ & $5.5 \pm 0.3$&  -- \\
$20344$ & 2018 Apr 20 & $1820$& $0.05\pm 0.01  $& $5.4 \pm 0.4$ & $5.5 \pm 0.4$&  -- \\
$20345$ & 2018 Apr 22 & $1822$& $0.05\pm 0.01  $& $4.4 \pm 0.4$ & $5.5 \pm 0.4$&  -- \\
$20346$ & 2018 Apr 24 & $1824$& $0.05\pm 0.01  $& $4.3 \pm 0.4$ & $5.5 \pm 0.5$&  -- \\
$20347$ & 2018 Apr 25 & $1825$& $0.06\pm 0.01  $& $5.1 \pm 0.4$ & $5.5 \pm 0.4$&  -- \\ \hline
\end{tabular} 
\caption{Time since the magnetar's outburst (2013 April 25 from \citet{kennea2013swift}), count rate $\left \langle \epsilon \right \rangle \times Q_{\text{mag}}$ from the magnetar at the position of Sgr A*'s extraction region, detected quiescent count rate $Q_{\text{eff}}$, predicted quiescent count rate $Q_{\text{Nowak}}$ according to \citet{nowak2012chandra} and $Q_{\text{sgr}}$ for each ObsID. From ObsID 18731 and forward in time, the magnetar is faint enough that count rates in the testing regions fall to background level, resulting in $\epsilon < 0$. This is why the corresponding $Q_{\text{sgr}}$ values are undefined since it implies that $\left \langle \epsilon \right \rangle \times Q_{\text{mag}}$ is negligible such that $Q_{\text{eff}} \approx Q_{\text{sgr}}$. 
\\\hspace{\textwidth} $a$: The apparent discrepancy between $Q_{\text{sgr}}$ and $Q_{\text{Nowak}}$ is due to $\epsilon$ from that ObsID being the highest, but the global value $\left \langle \epsilon \right \rangle$ is used when computing $Q_{\text{mag}}$ and $Q_{\text{sgr}}$ is thus overestimated. 
\\\hspace{\textwidth} $b$: A similar discrepancy occurs, but it is caused by a feature in the light curve that could be an undetected faint flare.}
\label{tab:Magn}
\end{table*}

\section{Flare detection}
\label{sec:BB}
\subsection{Bayesian Blocks and prior calibration}
\label{sec:BB_prior}
We choose Bayesian Blocks \citep{scargle2013studies} as our flare detection algorithm and use the implementation from \cite{williams2017pwkit} for $Chandra$ event files. The algorithm assumes that the data can be separated into different blocks, each with a constant count rate. If it detects a significant change in count rate at a specific point (called a change point) in the data, it will start a new block; where "significant" depends on the prior number of change points, $ncp\_prior$.

Since $ncp\_prior$ is linked to the expected number of blocks, it can be mathematically connected to the probability of a given change point being a false positive. We refer to this false alarm probability as $p_0$. The variable $ncp\_prior$ also depends on the number of events $N$ in a given light curve and on the underlying statistics of the data. The algorithm needs to be properly calibrated on signal-free data before being applied to real data. We modify \cite{williams2017pwkit}'s implementation to take this into account. See Appendix \ref{app:prior} for a detailed discussion of our calibration.

In this work, we adopt $p_0 = 0.05$. Since flares are made of at least 2 change points, this translates to a false positive rate of $p^{2}_{0} = 0.25\%$ (except for the case of edge flares, which are made up of only 1 change point leading to a false positive rate of 5\%).

For detections of individual events, we assume Poisson errors on our blocks (instead of the bootstrap simulations in the implementation of \citet{williams2017pwkit}).

\subsection{Pile-up}
\label{sec:pileup}
When two or more photons hit the same detector region in less than 1 frame time, they are miscounted as a single event with an energy corresponding to the total energy of the photons. This phenomenon is called pile-up and primarily affects bright point sources on the CCD. Dispersed photons can also be piled up, but much higher count rates are required because the flux is spread over more pixels; for the low brightness of Sgr A* the first-order events are effectively free of pile-up.

Pile-up can be described by Equation 2 of \citet{nowak2012chandra}:

\begin{equation}
\label{eq:crpiled}
    f_d = 1 - \frac{\left[ e^{\alpha \Lambda_i}-1 \right]e^{-\Lambda_i}}{\alpha \Lambda_i}
\end{equation}
where $\Lambda_i$ is the incoming (unpiled) counts per frame, $f_d$ is the fraction of events lost due to pile-up and $\alpha$ is the grade migration parameter (representing the fraction of recorded piled events) which we assume to be $\alpha = 1$ as shown in Equation 2 and Figure 2 of \citet{nowak2012chandra}.

For gratings data (frame time of 3.14 s), we unpile the flaring count rates in the zeroth order of each flaring block. We retrieve an average 0th/1st order flaring count rate ratio across all XVP flares of $\sim$ 1.6, consistent with \citet{nowak2012chandra}. 

\subsection{X-ray flare energies}
\label{sec:crtoenergies}
To accurately compare between grating and non-grating flare energies, we must convert counts to energy. Since Sgr A*'s flares have been shown to have very similar X-ray spectra \citep{nowak2012chandra,neilsen2013chandra}, their count rates are directly proportional to their fluxes given an ISM absorption model. We use \texttt{XSPEC} \citep{arnaud1996xspec} with a model of \texttt{dustscat*tbabs*powerlaw}, the abundances of \citet{wilms2000absorption} and the cross sections of \citet{verner1996atomic}. We take a power-law index of $\Gamma=2$, an hydrogen column density of $N_H = 14.3 \times 10^{22}$ cm$^{-2}$, a dust scattering optical depth $\tau = 0.324(N_H/10^{22} \text{ cm}^{-2})$ and we normalize to the the brightest flare of 2012, which has an absorbed 2-8 keV flux of $F^{\text{abs}}_{\text{2-8}} =8.5^{+0.9}_{-0.9} \times 10^{-12} \text{ erg}\text{ }\text{cm}^{-2}\text{ s}^{-1}$ and an unabsorbed 2-10 keV luminosity of $L^{\text{unabs}}_{2-10} = 19.2^{+7.2}_{-3.7} \times 10^{34} \text{ erg}\text{ s}^{-1}$ \citep{nowak2012chandra}. The unabsorbed emitted energies reported in Tables \ref{tab:XVP-flares} and \ref{tab:Post-XVP-flares} are pile-up corrected and quiescence subtracted.

With this model, we find a conversion factor between unpiled, quiescence subtracted 2-8 keV flare count rates and their unabsorbed 2-10 keV luminosity of $0.0077 \text{ ct}/$(10$^{34}$ erg) for gratings flares and $0.013 \text{ ct}/$(10$^{34}$ erg) for subarray flares. The difference comes from the effective areas of these instrument modes. Indeed, ACIS-S/HETG 0th+1st orders have an effective area of $\sim$ 200 cm$^{2}$ whereas ACIS-S has an effective area of $\sim$ 300 cm$^{2}$ (at 5 keV\footnote{\url{http://cxc.harvard.edu/caldb/prop_plan/pimms/}}).

\subsection{Blocks classification}
\label{sec:BB_class}

\begin{figure}[h]
    \centering
    \includegraphics[width=0.47\textwidth]{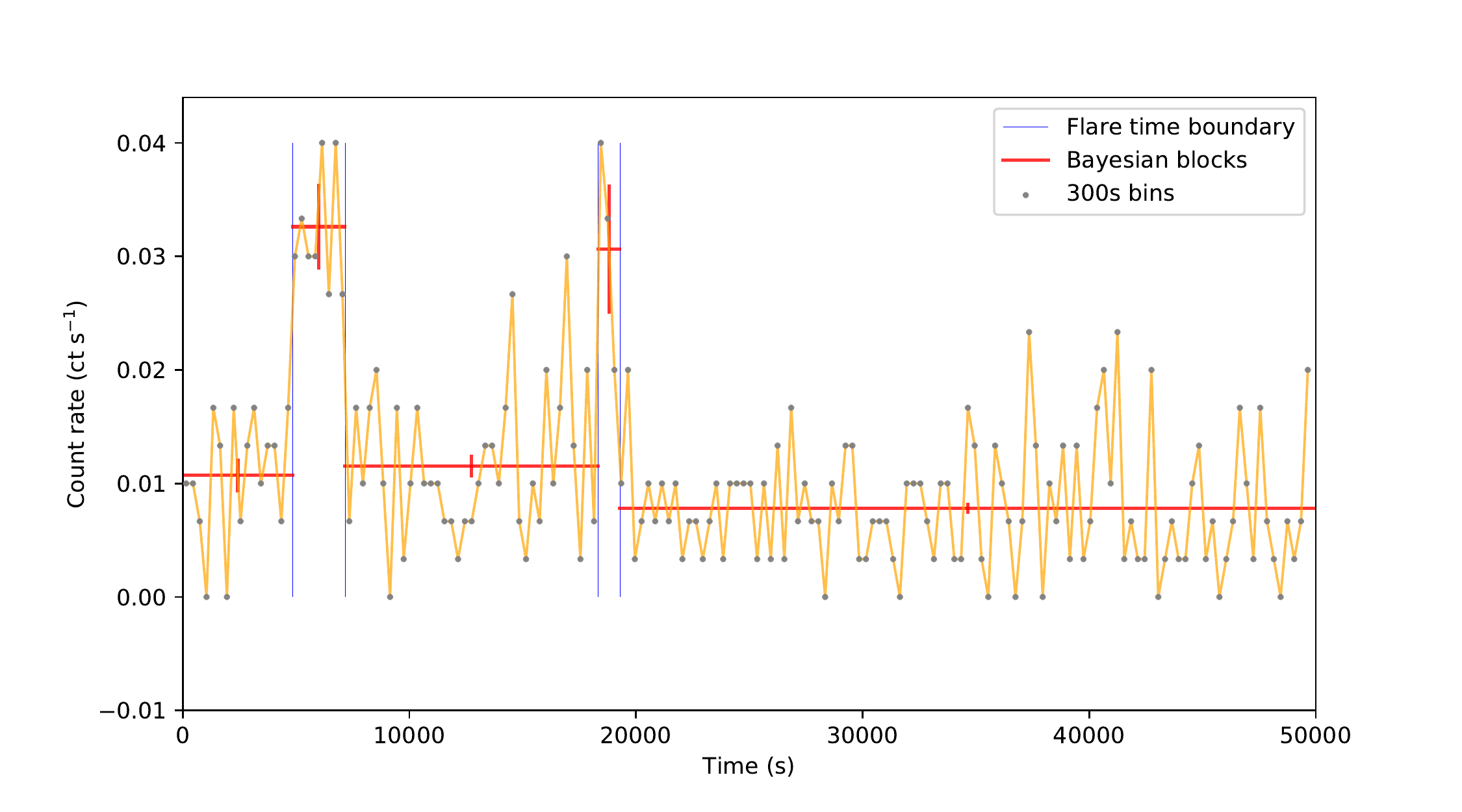}
    \caption{Light curve of ObsID 15045 (2013 October 28). The bin time is 300 s, the Bayesian Blocks are in red with their associated Poisson errors, the rightmost block is the quiescent block and the blue vertical bars indicate the beginning and end times of the two detected flares with the criterion explained in Section \ref{sec:BB_class}. If we relax the constraint, the block between the two flares will be considered a flaring block as well, thus leading to the detection of a single flare instead of 2.}
    \label{fig:15045}
\end{figure}

In addition to the Bayesian Blocks implementation, we define a set of criteria to determine which blocks are flaring blocks. This is especially important since we perform extensive Monte Carlo simulations and need to treat simulations and observations consistently. Since Sgr A* has a low flaring rate \citep[$\sim$ 1.1 flare day$^{\text{-1}}$;][]{neilsen2013chandra}, its quiescent count rate can be determined from the longest block in each observation. Given that the uncertainty on that block's count rate will be rather small (since it spans a long period), we consider a block as a flaring block if the difference between its count rate and the quiescent count rate is greater than $3\left(\sigma_{Q}+\sigma_{block}\right)$, where $\sigma_{Q}$ and $\sigma_{block}$ are the quiescent and flaring block's Poisson error on their respective count rate. We tried adding the errors in quadrature or using less than $3\left(\sigma_{Q}+\sigma_{block}\right)$ but found that this criterion behaves the best in general. That is, not adding the errors in quadrature increases the relative importance of the quiescent block's uncertainty (since it will always be the smallest), reducing the chance of having other quiescent-like blocks (also long with relatively low count rates) considered flares. Another way of seeing this is that the error on the actual quiescent count rate is likely larger than the error on the longest block. For blocks significantly above quiescence (which thus have larger error bars), adding uncertainties in quadrature instead makes a minimal difference.  Figure \ref{fig:15045} displays two distinct flares that would be grouped together into one flare by using less than $3\sigma$ or by using $3\sigma$ but adding the errors in quadrature (the block between them would be considered a flare).

Once all the flaring blocks have been found, we unpile them. We then multiply each block's count rate by its duration, add the resulting number of counts from each flaring block together, and divide it by the total duration of all the blocks in the flare, resulting in the flare's average count rate. We then subtract the quiescent count rate, and convert the resulting number of counts to an energy using the conversion factors presented in Section \ref{sec:crtoenergies}.

\begin{figure}[h!]
    \centering
    \includegraphics[width=0.47\textwidth]{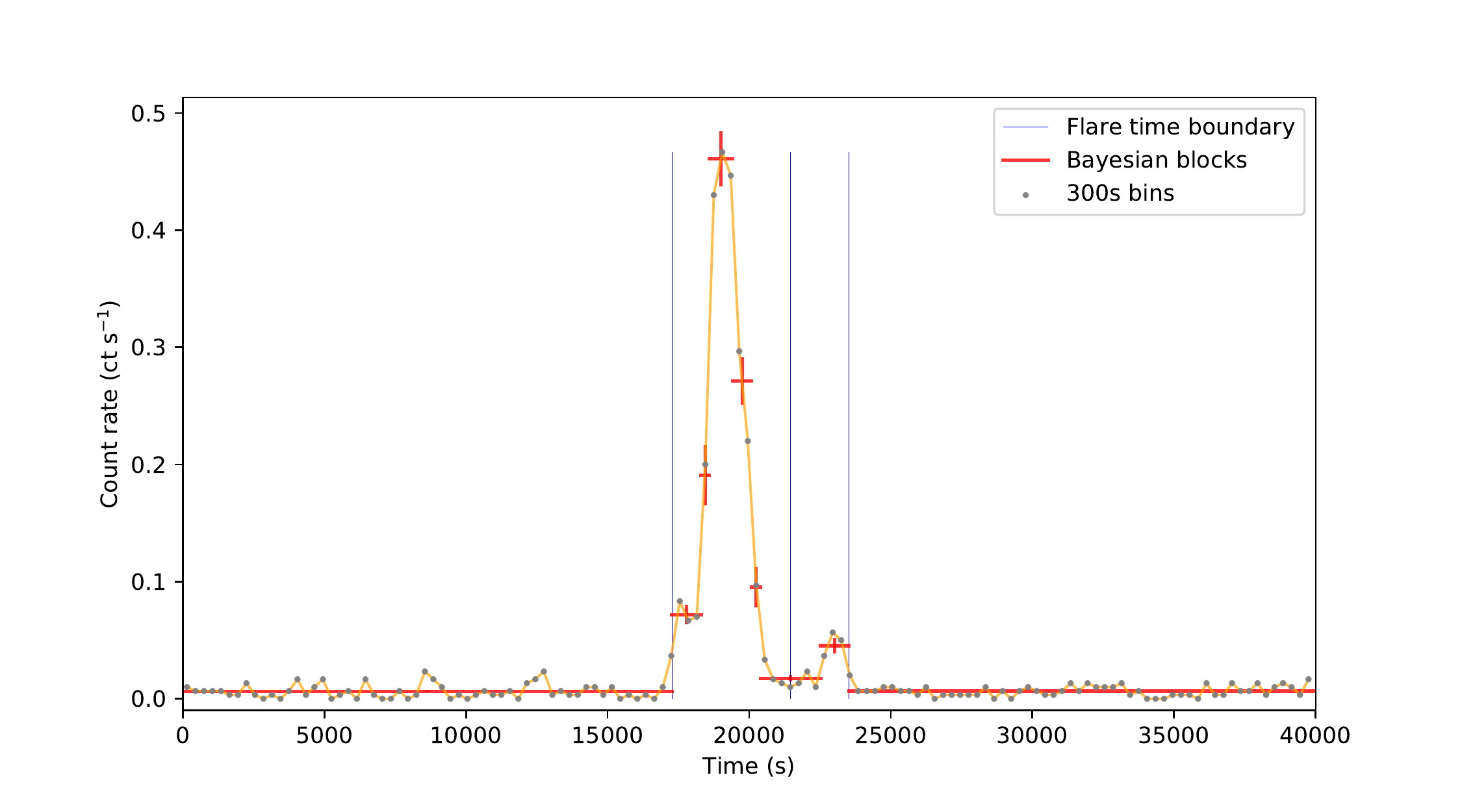}
    \caption{Like Figure \ref{fig:15045} but for ObsID 16218 (2014 October 20). There is an apparent dip between two flares. However, the block between the two flares is still significantly above quiescence (from the $3\sigma$ criterion). If we use the flare separation criterion, this block is cut in half and we end up with two flares instead of one.}
    \label{fig:16218}
\end{figure}

It is possible in principle for two long and bright flares to occur too close together in time to be distinguishable. This is discussed by \citet{yuan2015systematic}, where the authors find that such a phenomenon is quite unlikely (only 1/3 of their flares that show evidence of substructures have a probability above 5\% of being made of multiple flares). We introduce another criterion (referred to as the flare separation criterion from now on) in an effort to distinguish close flares and test our whole analysis with and without it. The criterion stipulates that if there is a series of flaring blocks with one of them (not the first nor the last one) having a count rate significantly lower than the others (the count rate difference is larger than 3 error bars added in quadrature\footnote{Similarly to the $3\sigma$ criterion, we found this by trial and error and settled on this formulation after determining it to be the best behaving in general.} between the block and its neighbours), then it is flagged and the flare is considered to be made of two flares. This block is separated in half, with its first half being associated with the left flare and the other half with the right flare. This occurs in only 3 observations, all of which are Post-XVP (ObsIDs 15043, 16218 and 20346). In Section \ref{sec:simulation}, we do not use this criterion unless specified otherwise. Figure \ref{fig:16218} shows ObsID 16218 as an example. We report our flares in Tables \ref{tab:XVP-flares} and \ref{tab:Post-XVP-flares}. 

\subsection{Detected flares}
\label{sec:detflares}
We find that the flaring rate is consistent across the datasets as we detect 40 XVP flares in 3 Ms ($1.2 \pm 0.2$ flare day$^{\text{-1}}$) and 18 Post-XVP flares in 1.56 Ms ($1.0 \pm 0.2$ flare day$^{\text{-1}}$). Light curves are available\dataset[here]{\doi{10.5281/zenodo.3373989}}. We compute the XVP and Post-XVP observed differential flare energy distributions dN/dE (see Figure \ref{fig:XVPvsPostXVP}). The upper and lower $1\sigma$ error bars on each data bin are computed with the inverse of the Poisson Cumulative Distribution Function (CDF) for the 15.87th and 84.13th percentiles, respectively, divided by the width of that bin. If we add the flare separation criterion, we identify the same number of XVP flares, but we find 21 Post-XVP flares giving a rate of $1.2 \pm 0.3$ flare day$^{\text{-1}}$; all are consistent with \cite{neilsen2013chandra}. The properties of each flare are provided in Tables \ref{tab:XVP-flares} and \ref{tab:Post-XVP-flares}. We compare our detected flares in detail with other works \citep{neilsen2013chandra,ponti2015fifteen,mossoux2017sixteen} in Appendix \ref{sec:compar}. Our results are mostly consistent with the previous authors, but some differences occur for the faint/short flares due to different criteria for flaring blocks, the Bayesian Blocks calibration and the flare detection method.

\section{Simulations}
\label{sec:simulation}
To search for significant changes in the flaring behaviour of Sgr A*, we present a model for X-ray flares based on a Monte Carlo approach. To compute their associated CDF, the model assumes Gaussian flares and that the flare energy and duration distributions follow a power-law.

For each ObsID in a given instrument mode, the model generates an event list which requires a quiescent count rate. For XVP HETG data, since we do not detect any trend over long periods in the observed quiescent count rates (see Table \ref{tab:XVP-flares}), we simulate a random Poisson count rate around the median of the observed values, 0.0063 ct s$^{\text{-1}}$, for each simulated observation. For Post-XVP light curves, we use the analysis in Section \ref{sec:magn}; we draw random values around $\left \langle \epsilon \right \rangle \times Q_{\text{mag}}$ and $Q_\text{sgr}$ for the corresponding ObsID and add them together. For simulated light curves from 2016 July 12 and on, we draw random Poisson quiescent count rates from $Q_\text{sgr}$.

The total simulated time for a given ObsID is given by the sum of the duration of each of its Bayesian Blocks (given in Tables \ref{tab:XVP-flares} and \ref{tab:Post-XVP-flares} as the exposure) plus an additional 10 ks before and after to accommodate edge flares in the analysis. We conservatively choose 10 ks to be positive to pick up any simulated flare since our longest flares have $\sigma$ = 2 ks (see Section \ref{subsec:mod-para}). Flare times are placed randomly in the simulated observation according to a given Poisson flaring rate. 

For each flare time, a duration and an emitted energy are randomly drawn from their respective CDFs (assumed to be CDFs of power-law distributions which might differ from the observed ones; see Section \ref{subsec:mod-para}). Like \citet{neilsen2013chandra}, we define the standard deviation $\sigma$ of a Gaussian flare as its duration divided by 4 (we have tried other values than 4, but, using simulated flares, we find that 4 is the most reliable value for typical flare parameters). Given the energy and duration of a flare, we compute its mean luminosity and convert it to a count rate via the scaling established in Section \ref{sec:crtoenergies}. The count rate amplitude $A$ of the Gaussian flare is also determined. The flare is then piled and the final event list generated after removing the additional 10 ks that were added before and after the simulated ObsID. To generate event lists from the Gaussian flare and the quiescent count rate\footnote{Note that it is also possible to have zero or multiple flares in a simulated observation.}, we follow Appendix D of \citet{mossoux2017sixteen}. We compute the expected number of counts within the simulated observation, draw a random Poisson number around it and calculate the light curve's CDF (CDF of a constant rate plus a Gaussian) which is used to assign a random time to each event. For explanations regarding how pile-up is handled in simulated gratings light curves, see Appendix \ref{sec:App_puGrating}.

We run Bayesian Blocks on the simulated event lists and retrieve the detected flares in the same way we did for the data in Section \ref{sec:BB}. (See Figure \ref{fig:sim14392} for an example of a simulation that uses the detected parameters of the two flares from ObsID 14392 as input. We also provide a code that simulates flares of given duration and energy for each instrument mode.\footnote{A code that simulates flares of given duration and energy for each instrument mode is available at \url{https://github.com/Elie23/X-ray-flare-simulator}}) We do this for each dataset and reconstruct the flare energy distribution to compare it with observations. We split the simulation data into the same logarithmically spaced energy bins as the data to compare them. We simulate each dataset 3000 times and produce $15\%$ - $85\%$ ($70\%$ intervals) confidence intervals for each bin. By using the same model parameters for the XVP and post-XVP datasets, we are able to test whether they are consistent with a single flare distribution. If this model matches both datasets (as determined from the simulated $70\%$ confidence intervals above), then there is likely no compelling evidence for a change in Sgr A*'s X-ray flaring behavior between those datasets. Instead of finding change points, we fail to reject the null hypothesis, i.e, we find that there exists a selection of parameters that match the data before and after each potential change point (see Section \ref{subsec:mod-para} below). Figures \ref{fig:XVPvsPostXVP} and \ref{fig:G2ChangePoints} compare the XVP and Post-XVP datasets as well as the Pre and Post-G2 datasets for different potential change points in time (after 2014 April 4th and 2014 August 30th, respectively). The confidence intervals are obtained from a single set of model parameters described below.

\subsection{Simulations of the XVP and Post-XVP observations}
\label{subsec:mod-para}
\begin{figure}[h!]
    \centering
    \includegraphics[width=0.47\textwidth]{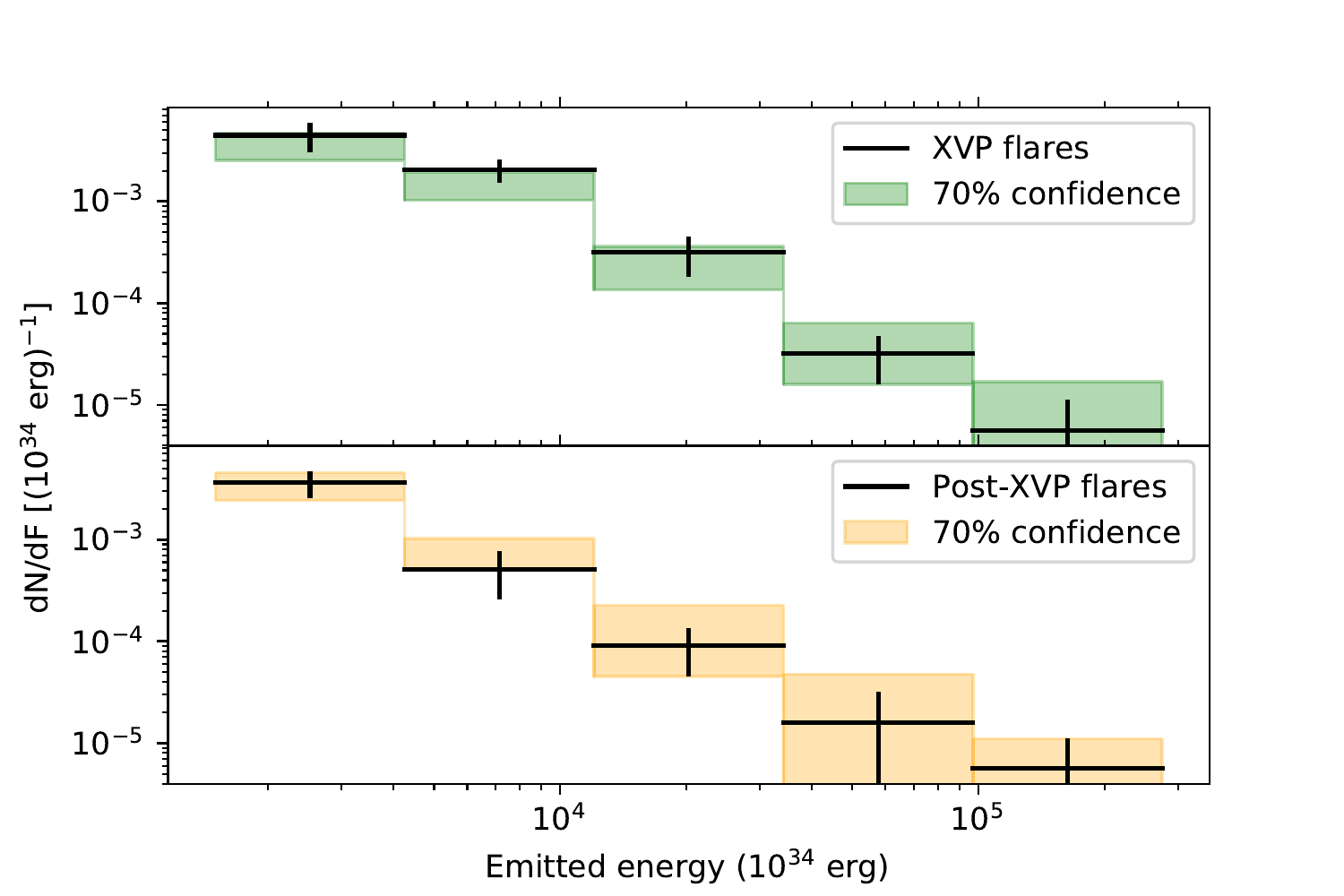} 
    \caption{Observed binned unabsorbed energy differential distribution of XVP and Post-XVP (black lines) flares (see Tables \ref{tab:XVP-flares} and \ref{tab:Post-XVP-flares}), with Poisson errors on the number of flares in that bin divided by its width. The colored shaded regions are the associated $70\%$ confidence regions obtained from 3000 Monte Carlo simulations.}
    \label{fig:XVPvsPostXVP}
\end{figure}

\begin{figure*}[t!]
    \centering
    \includegraphics[width=\textwidth]{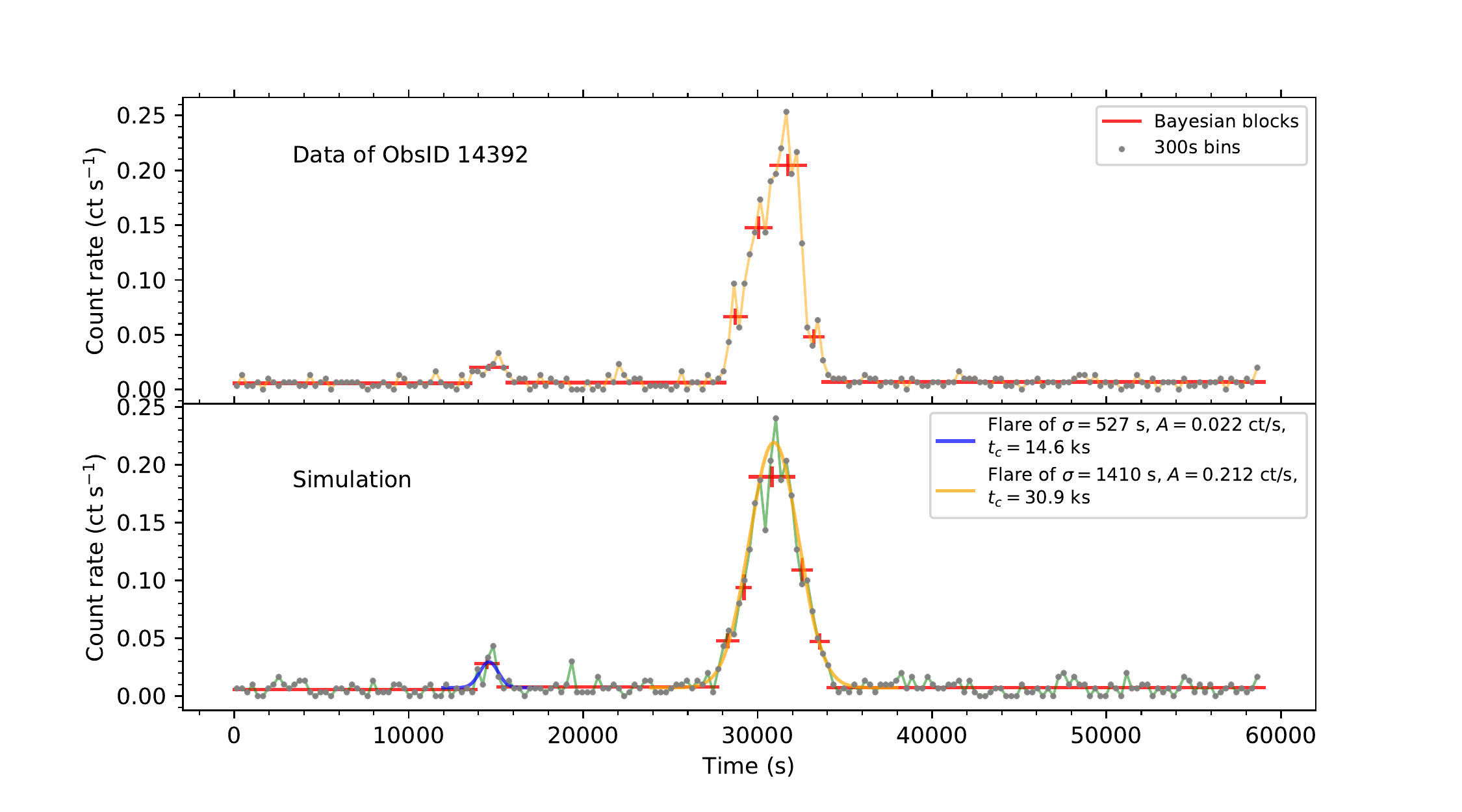}
    \caption{Top: 2-8 keV light curve of ObsID 14392 (combined zeroth and first order events) in 300s bins. Shown in red are the Bayesian Blocks. $Bottom$: Simulated light curve using the same quiescence and flare parameters as those measured in ObsID 14392 as input. The two flares detected in ObsID 14392 are represented by the blue and orange Gaussians, respectively. Their parameters are stated in the legend. The Bayesian Blocks retrieve flare energies of $3.5 \times 10^{37}$ erg and $100.0 \times 10^{37}$ erg compared to the input values of $3.74 \times 10^{37}$ erg and $101.1 \times 10^{37}$ erg.}
    \label{fig:sim14392}
\end{figure*}

Since the XVP and Post-XVP datasets were obtained in different instrument modes with different sensitivities, we might expect different flare energy distributions. We can test if such a change is physical or due to a change in instrument mode since our model takes both into consideration automatically. In Figure \ref{fig:XVPvsPostXVP}, we show the energy distributions from our XVP (top) and Post-XVP (bottom) datasets over-plotted with $70\%$ confidence regions produced by 3000 Monte Carlo simulations of each dataset. We used the following parameters (found by trial and error\footnote{Parameters that produce better agreement might exist.}): duration (energy) distribution power-law index $\Gamma_\text{Dura} = -0.8$ ($\Gamma_\text{Energy} = -1.7$), simulated duration (energy) range of 500 s to 8000 s ($1.3 \times 10^{37}$ erg to $275\times 10^{37}$ erg\footnote{From the most energetic flare in ObsID 15043.}) and a flaring rate of 52 flares per 3 Ms ($\sim 1.5$ per day). These may not represent the true physical parameters of Sgr A*, but they are sufficient for our test here. Since these parameters produce $70\%$ confidence intervals consistent with observations for both datasets, we infer that the flare distributions are consistent with each other. Many complex factors contribute to the wideness of the confidence intervals, namely the random draw from a CDF of a low number of flares, edge effects, overlapping flares, the inherent Poisson counting noise, and finally the Bayesian Blocks detection process itself. The duration and energy power-law indexes are in agreement with those reported by \citet{neilsen2013chandra}, but our flaring rate is higher  because it represents the intrinsic flare rate rather than the observed flare rate. This analysis effectively demonstrates that the XVP and Post-XVP flare distributions are consistent with each other; we discuss some limitations of this method in Section \ref{sec:Discus}.

If a simulated flare has a detected duration (or energy) outside the simulated parameter range, it is appended to the closest histogram bin. For XVP HETG simulations, we find no flare fainter than $1.3 \times 10^{37}$ erg because the detection efficiency rapidly drops below $\sim 2 \times 10^{37}$ erg, but we do find at least 1 flare brighter than $275\times 10^{37}$ erg in $8.8\%$ of our simulations. For Post-XVP simulations, we find at least 1 flare fainter (brighter) than $1.3 \times 10^{37}$ erg ($275\times 10^{37}$ erg) in $47\%$ ($2.3\%$) of simulations. This instrument mode has a much better sensitivity, such that its detection efficiency only starts to drop for energies below $\sim 1.5 \times 10^{37}$ erg. Bayesian Blocks tend to underestimate energies of simulated faint flares because they are more likely to be detected for shorter periods ($< 4\sigma$). Also, Poisson noise causes simulated flares to contain less counts than their expected value half the time, automatically creating flares outside the simulated energy range. The reason behind bright flares being sometimes detected above the simulated energy range, besides Poisson noise, is that our algorithm tends to overestimate the energy within the brightest flares. There are two reasons behind this. First, simulated bright flares tend to be detected for longer by the Bayesian Blocks since a greater fraction of their Gaussian will be significantly above quiescence. Because we assume that all the flare's energy is stored within a duration of $4\sigma$ to compute its mean count rate to simulate it, this leads to an overestimated energy when the Bayesian Blocks detect the flare for longer than that. However, since less than $5\%$ of a Gaussian's area is outside this $4\sigma$ range, the degree of this overestimation is not concerning. Second, when we pile the flare before creating the event list, we pile only the mean count rate to conserve the shape of the Gaussian which is needed for the CDF and thus the creation of the event list itself. When we unpile, we proceed the same way as for actual observations; we unpile each flaring block. These two effects combined create an overestimation of only $\sim$ 10-15\% for simulated flares of $250\times 10^{37}$ erg. For flares of $\leq 100\times 10^{37}$ erg, these effects become negligible. Given the size of the flare energy bins, this is good enough for this work. We observe that 90\% of the flares brighter than $275\times 10^{37}$ erg have energies below $300\times 10^{37}$ erg; the rest are caused by rare overlapping flares. The percentage of simulations containing at least 1 detected energy greater than the simulated range is lower for Post-XVP simulations than for XVP simulations since the total XVP exposure is longer and the average exposure of each XVP observation is longer.

If we use the flare separation criterion (last paragraph of Section \ref{sec:BB_class}), we change the maximum energy to $150\times 10^{37}$ erg since the flare in ObsID 15043 becomes two flares. Leaving the other model parameters at their original values, we again find that the two datasets are consistent.

\subsection{Looking for change points around G2's pericenter passage}

\citet{ponti2015fifteen} and \citet{mossoux2017sixteen} report an increase in the bright flaring rate on August 31, 2014 which they argue is caused by the pericentre passage of G2 around Sgr A*. \citet{ponti2015fifteen} also argue that this change could be caused by a noise process, where the flaring rate is constant on average but shows clustering on shorter timescales which was only detected because of the increased monitoring frequency around G2's pericentre passage. These authors also mention that they do not observe this change if they limit themselves to $Chandra$ data only. Indeed, most (4 out of 5) of the bright/very bright flares responsible for this shift are from \textit{XMM}\textit{-}\textit{Newton}. \citet{mossoux2017sixteen} revisit this analysis, including two more recent bright $Chandra$ flares from ObsIDs 16966 and 17857 and a $Swift$ flare from February 2015.

We test if inclusion of new $Chandra$ data and our simulations retrieve such a change point. We explore four different potential change points; ObsIDs 16212 (2014 April 4th), 16214 (2014 May 20th), 16215 (2014 July 16th) and ObsID 16217 (2014 August 30th). The tests around April, May and July 2014 are motivated by the different G2 pericenter times from models. A purely Keplerian orbit gives an estimated time between the end of February to mid April 2014. The addition of a drag force pushes this date to between the end of May and mid-July, and if an inflow is also added, this prediction shifts to between July and September  \citep{gillessen2013pericenter,madigan2016using}. 

\begin{figure*}[t!]
    \centering
    \includegraphics[width=\textwidth]{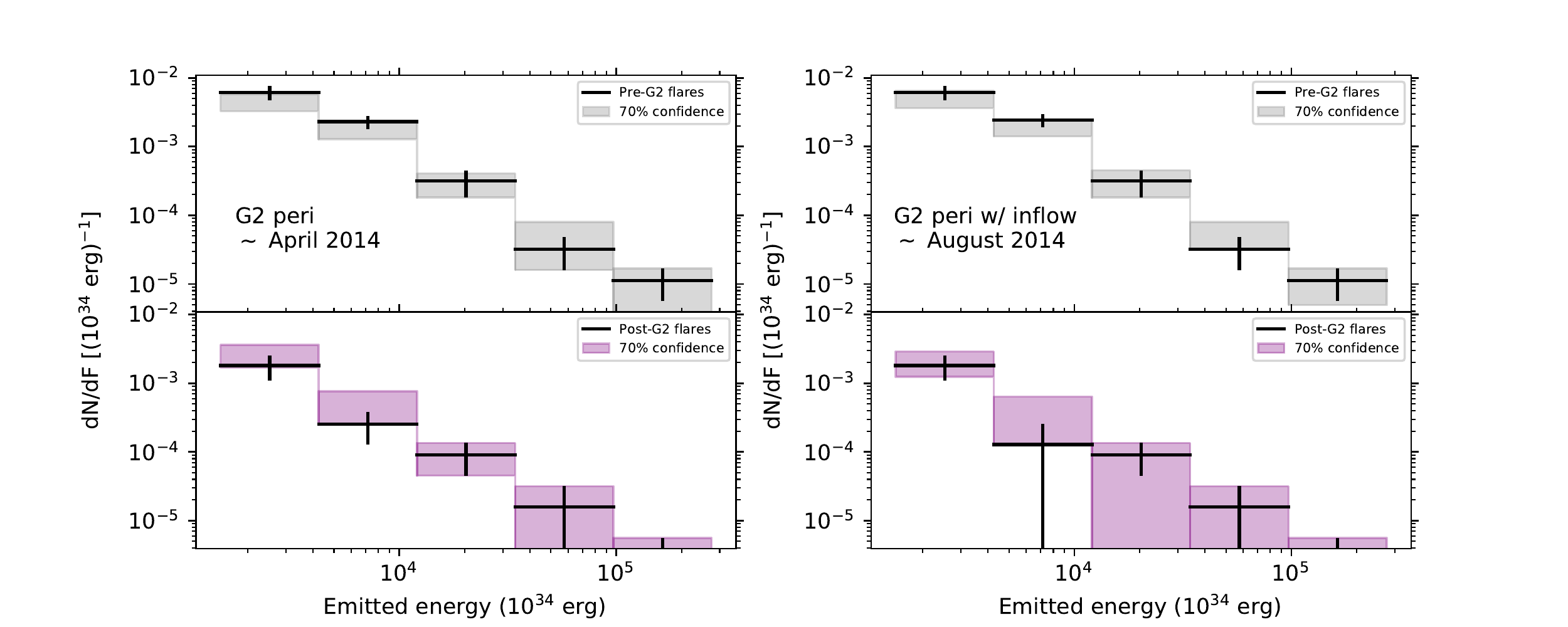}
    \caption{Similar to Figure \ref{fig:XVPvsPostXVP}, but for different change points around the pericenter passage of G2. The shaded regions also represent $70\%$ confidence regions obtained from 3000 Monte Carlo simulations. $Left$: Test for a change point after ObsID 16212 (2014 April 4th). $Right$: Test for a change point after ObsID 16217 (2014 August 30th).}
    \label{fig:G2ChangePoints}
\end{figure*}

We find that the same model parameters presented above produce $70\%$ confidence intervals consistent with observations for both datasets, once again failing to reject the null hypothesis and implying no change in Sgr A*'s X-ray flaring behaviour. Since the only ObsID in which we find flares in these ObsIDs is 16217, the only variation in the other ObsIDs is how the exposure is split between Pre and Post-G2. Therefore, we only show results for ObsIDs 16212 and 16217 in Figure \ref{fig:G2ChangePoints}. We observe that despite the removal of the flare at ObsID 16217 from the Post-G2 dataset, the entire dataset remains consistent with the Pre-G2 observations. We also test the change point (with and without the flare separation criterion) reported by \citet{mossoux2017sixteen} where they find a decrease in the flaring rate of the less energetic flares on 2013 July 27 (ObsID 15041). Similarly, \citet{ponti2015fifteen} report a tentative decrease in the moderate-bright flaring rate after 2013 June 5 (ObsID 14703) using their \textit{Chandra}-only dataset (see Section \ref{sec:Discus}). We do not recover either of these findings.

\section{Discussion}
\label{sec:Discus}

Based on the simulations and analysis presented here, we do not identify any significant change in the distribution of emitted energies of flares from Sgr A* at any point during the $Chandra$ monitoring campaign between 2012-2018. This is in contrast with the findings of \citet{ponti2015fifteen} and \citet{mossoux2017sixteen} who find that an increase in the bright flaring rate occurs around 2014 August 31 (after ObsID 16217); the latter also report a decrease of the faint flaring rate on 2013 July 27 (ObsID 15041); again, we do not find evidence for such changes in the emitted energy distributions. However, these authors also use data from \textit{XMM}\textit{-}\textit{Newton} and $Swift$\footnote{Using their \textit{Chandra} observations only, \citet{ponti2015fifteen} further find a slight decrease in the moderate-bright flaring rate following 2013 June 5. Our analysis does not detect such a signal. This discrepancy is likely due to its relatively low significance (96\%) and our simulations taking into account several detection biases such as different instrument modes.}, which cannot detect weak and moderate flares and most of their brightest flares are detected by \textit{XMM}\textit{-}\textit{Newton}. Focusing on the $Chandra$ data alone allows a systematic characterization of flares at all intensities, which motivates our choice to analyze data from this single observatory in the present work. 

In comparison to these similar analyses by \citet{ponti2015fifteen} and \citet{mossoux2017sixteen}, the present work is an improvement in several areas.

First, we perform a systematic analysis of flares of all intensities using a dataset with relatively uniform sensitivity.

Second, we calibrate our Bayesian Blocks algorithm using signal-free light curves, ensuring that our flare detection and significance estimates are accurate.

Third, we calculate the flux of each flare assuming a constant conversion between count rate and emitted energy \citep[see also][]{neilsen2013chandra,yuan2017systematic}. This avoids difficulties fitting flare spectra with poor statistics \citep[eg.,][]{mossoux2017sixteen}.

Fourth, we account for any dependence on the instrument mode in our simulations. Our simulations also use an instrument mode dependent pile-up scheme and produce entire renderings of each dataset, taking into account automatically the relative flare detection efficiencies of each dataset. This model is used to test whether the same model parameters can successfully ---as we find in this work--- reproduce the observed flare energy distributions of different datasets, then those datasets are consistent.

However, the present work shares some shortcomings with previous studies. Namely, we assume that the flares occur at random Poisson times. However, it's possible that Sgr A*'s flaring behaviour exhibits clustering. \citet{yuan2015systematic} show that flare clustering on timescales of 20-70 ks is significant at the $96\%$ level (for their gratings dataset, but for their ACIS-I sample this number drops to $\sim 50\%$) and may be described by a piecewise-deterministic Markov process \citep{davis1984piecewise}. Furthermore, we assume that flare durations and energies are independent. However, \citet{neilsen2013chandra} show a moderate correlation ($\rho \sim 0.54$) between the two. Our simulations \textit{a-posteriori} recover some of this correlation ($\rho \sim 0.23$) from detection biases. We consider this sufficient as \cite{yuan2017systematic} report no correlation between those quantities.

Despite these potential drawbacks, based on our analysis of the emitted energy distributions we believe the change points reported by \citet{ponti2015fifteen} and \citet{mossoux2017sixteen} are unlikely to be indicative of an increase in Sgr A*'s bright flaring activity caused by G2. If the increased bright flaring rate observed by these authors was due to G2, this process should stay active for at least a viscous timescale, $\sim$ 3-10 years at $\sim 2000$ R$_s$ \citep[][but see \citet{do2019unprecedented}]{yuan2014hot,ponti2015fifteen,mossoux2016multiwavelength} assuming the canonical viscous parameter value $\alpha=0.1$, with $\alpha$ as defined as in \citet{shakura1973black}. In this scenario, our model should have detected this persistent signal with the additional $Chandra$ exposure of 652 ks across 14 observations from June 2016 to April 2018 (from ObsID 18731 to 20347). In addition to the previously tested change points, it could be possible that the bright flaring rate was delayed and started increasing from those newer observations. This delayed timescale of a few years after pericentre probes the estimated timescale before an increased activity begins according to RIAF models \citep{yuan2014hot} and is worth testing. We observe 3 bright flares between June 2016 and April 2018 \citep[according to the definition of > 120 counts from][corresponding to an unabsorbed energy of 9.2 $\times 10^{37}$ erg in our case]{ponti2015fifteen}, translating to a bright flaring rate of $0.4 \pm 0.3$ flare day$^{\text{-1}}$. For the whole Post-XVP dataset, we detect 6 bright flares in 1.56 Ms giving a bright flaring rate of $0.3 \pm 0.1$ flare day$^{\text{-1}}$. For the XVP dataset, we have 10 bright flares in 3 Ms translating to a bright flaring rate of $0.29 \pm 0.09$ flare day$^{\text{-1}}$. All these rates are consistent within errors. The 3$\sigma$ upper limit of 3 is 9 giving a rate of $1.2$ flare day$^{\text{-1}}$, still lower than the predicted rate of $2.52 \pm 0.98$ flare day$^{\text{-1}}$ as found by \citet{ponti2015fifteen}. This increase is thus absent from our $Chandra$-only dataset. Interestingly, \citet{ponti2015fifteen} also don't find an increase in the bright X-ray flaring rate if they only use their $Chandra$ data. In fact, the increase in the bright flaring rate found by \citet{ponti2015fifteen} is mainly due to 4 bright \textit{XMM}\textit{-}\textit{Newton} flares detected in only 133 ks of observations at the end of 2014. The fact that this significant increase is not present in their (nor our) \textit{Chandra} sample could be explained by flare clustering as mentioned above \citep{yuan2015systematic,yuan2017systematic} and suggested by \citet{ponti2015fifteen} In any case, this apparent discrepancy shows the importance of coordinated observations between different X-ray observatories to improve our understanding of their relative flare detection efficiency. 

From the previously mentioned viscous timescale, it's also possible that G2 still hasn't caused an increase in Sgr A*'s X-ray flaring rate, but will in the near future. \citet{schartmann2012simulations,kawashima2017possible} also predict an increased activity years and 5-10 years, respectively, after G2 as mentioned in the introduction. Future observations of Sgr A* will shed light on this matter.

It is worth noting that another gas cloud, G1, which has nearly identical orbital parameters to G2 \citep{pfuhl2015galactic}, underwent a pericentre passage near Sgr A* in 2001 and no change in Sgr A*'s quiescence or flaring properties were observed \citep{yuan2015systematic}.

\section{Conclusion}
\label{sec:Conclu}

In this work, by using Bayesian Blocks and adding flaring blocks together with a systematic method, we detected and characterized 58 flares (40 XVP flares and 18 Post-XVP flares, respectively) in 4.5 Ms of $Chandra$ observations from 2012 to 2018. We simulated X-ray light curves built from parameters such as flaring rate, quiescent count rate, flare count rate to luminosity conversion factor, energy and duration distributions, exposure time and pile-up. By simulating these light curves to reproduce different datasets (XVP and Post-XVP or Pre-G2 and Post-G2) using the same parameters, we found no evidence of a change point in the energy distribution above 95\% confidence at any point in Summer or Fall 2014, failing to reject the null hypothesis. We have shown empirically that ($1.3\pm0.2$)\% of the count rate from the transient magnetar SGR J1745-2900 contaminates Sgr A*'s extraction region and quantified it as a function of time since its outburst on April 25 2013. Our findings indicate that the previously reported increased bright flaring rate in Summer 2014 \citep[e.g.,][]{ponti2015fifteen,mossoux2017sixteen} is absent from our $Chandra$-only data. Future $Chandra$ observations will allow us to determine if G2 caused a delayed increase in the flaring rate. Extraordinary $Chandra$ observations coordinated with EHT and GRAVITY of Sgr A* in the coming years will improve our understanding of Sgr A*'s X-ray flares.


\bigskip
We thank all the members of the Sgr A* Chandra XVP collaboration (\url{https://www.sgra-star.com/collaboration-members}), and we are immensely grateful to the Chandra scheduling, data processing, and archive teams for their support during these Sgr A* monitoring campaigns. We also thank Nicolas Cowan, Melanie Nynka, John Ruan, Hope Boyce, Peter Williams, Emmanuelle Mossoux and Jeffrey Scargle for useful conversations and suggestions that improved this manuscript. \'{E}.B. and D.H. acknowledge support from the Natural Sciences and Engineering Research Council of Canada (NSERC) Discovery Grant and the Fonds de recherche du Qu\'{e}bec--Nature et Technologies (FRQNT) Nouveaux Chercheurs and Bourse de ma\^itrise en recherche (B1X) programs. S.M. is thankful for support from an NWO (Netherlands Organisation for Scientific Research) VICI award, grant Nr. 639.043.513. D.H. acknowledges support from the Canadian Institute for Advanced Research (CIFAR). \'{E}.B. also gratefully acknowledges support from the McGill Space Institute Graduate Fellowship and the Lorne Trottier Accelerator Fellowship.

\appendix
\section{Bayesian Blocks prior calibration}
\label{app:prior}
To calibrate $ncp\_prior$ on Poisson noise, we follow the procedure suggested in \citet{scargle2013studies} and applied in Appendix A of \cite{mossoux2015study}. We simulate signal-free light curves by choosing a number of expected counts $N$ and a constant noise count rate $cr$. 

The resulting event list is sent to the Bayesian Blocks algorithm, which returns the number of change points detected. For each $N$ considered, we evaluate values of $ncp\_prior$ between 3 and 15, with a step of 0.1, and generate 1000 random Poisson light curves. $p_0$ is the sum of the detected change points divided by the number of light curves. We stick with $p_0 = 0.05$ and get a relation between $ncp\_prior$ and $N$.

\begin{figure}[h!]
    \centering
    \includegraphics[width=0.7\textwidth]{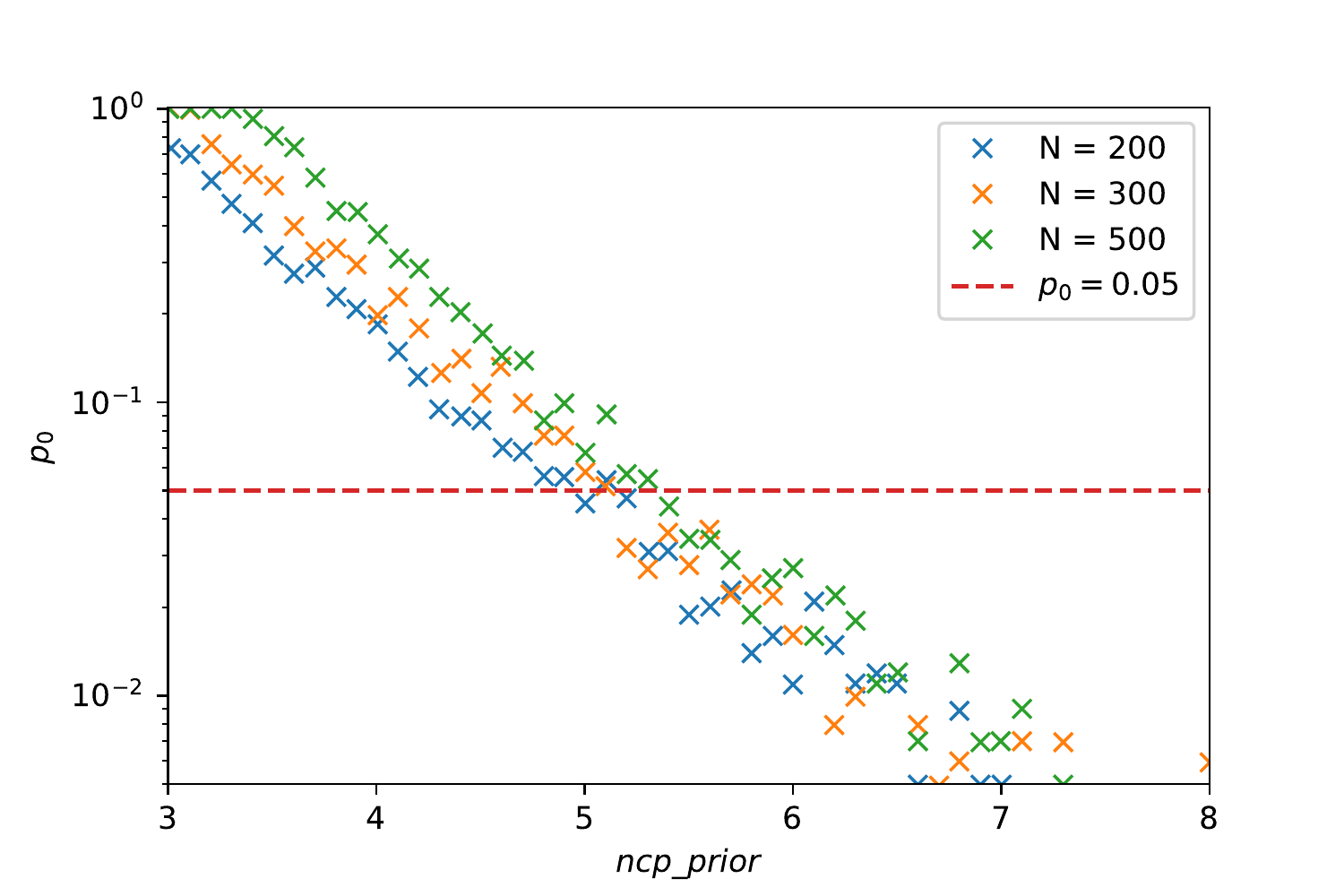}
    \caption{Number of change points detected in 1000 random Poisson-generated signal-free light curves divided by 1000 ($p_0$) as a function of $ncp\_prior$ for different number of expected counts $N$ in the light curves. The noise count rate is 0.01 ct s$^{\text{-1}}$. The red horizontal dotted line indicates $p_0 = 0.05$.} 
    \label{fig:noise_cr_3curves}
\end{figure}

\begin{figure}[h!]
    \centering
    \includegraphics[width=0.7\textwidth]{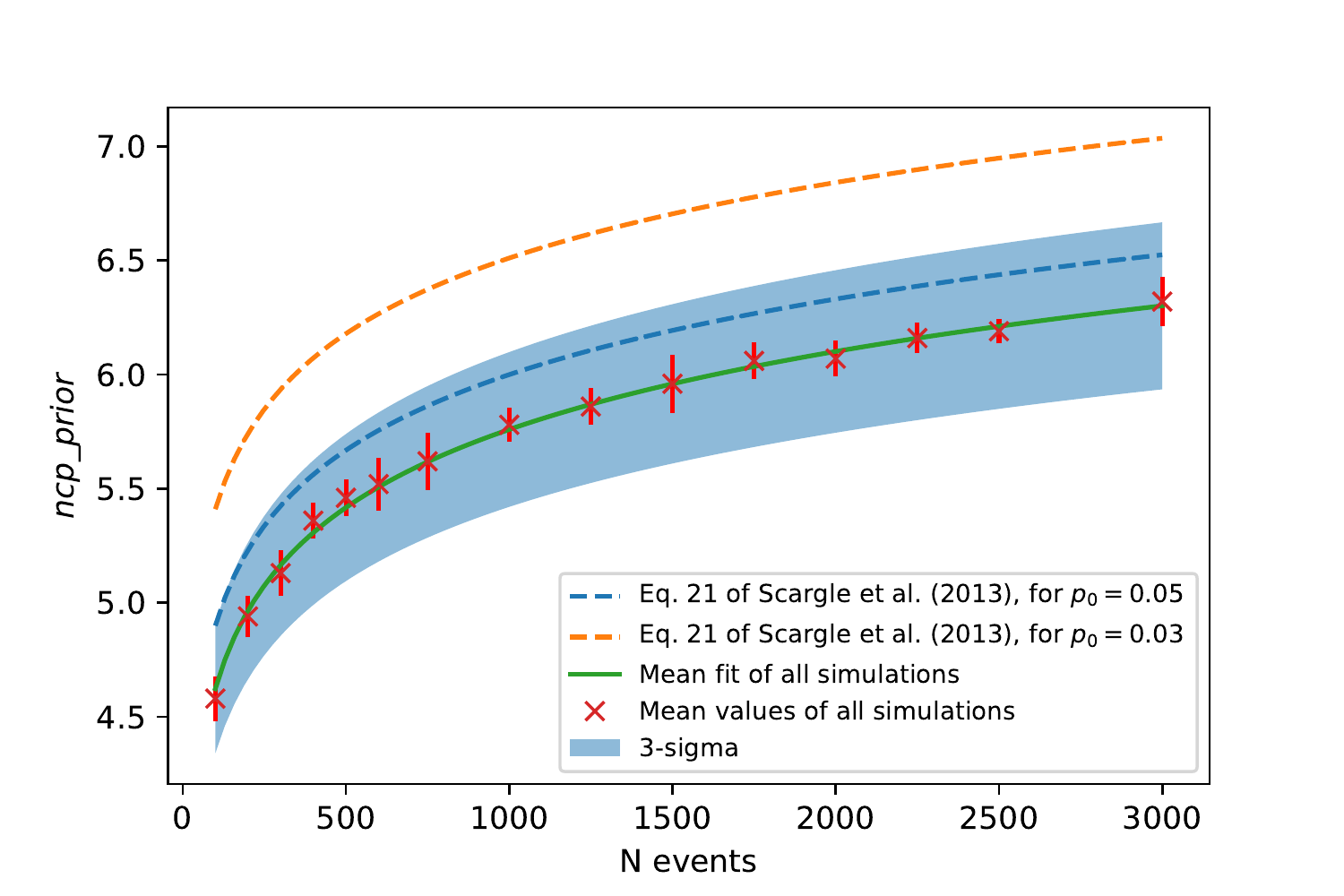}
    \caption{Summary plot showing the results of 10 different calibration runs for different noise count rates with $p_0 = 0.05$. Each red point represents the mean value of $ncp\_prior$ found across those runs for that given expected number of events $N$, and the error bars are their standard deviation. The green line is the resulting best fit and the blue shaded region is its $3\sigma$ error envelope. The blue and orange dotted lines are the calibrations obtained by \cite{scargle2013studies} for similar simulations for Gaussian noise for $p_0 = 0.05$ and $p_0 = 0.03$, respectively.}
    \label{fig:meanof10simu3sig}
\end{figure}

Since the value of $ncp\_prior$ chosen with our method is a bit noisy (e.g. Figure \ref{fig:noise_cr_3curves}), we run the simulation for 10 different noise count rates and take the mean of the different $ncp\_prior$ values obtained at each $N$, as well as their standard deviation (see Figure \ref{fig:meanof10simu3sig}). We choose different count rates to double-check our intuition that changing the count rates does not change the results. Indeed, if the count rate changes, the only modification in the Bayesian Blocks should be the mean count rate of each block (i.e., the probability of finding false change points shouldn't change). 

We fit the following function to the data points in Figure \ref{fig:meanof10simu3sig}:

\begin{equation}
\label{eq:ncp_prior_fit}
ncp\_prior(N) = A\log N + B
\end{equation}
where $ncp\_prior(N)$ is the fitted value of the prior for a given expected number of events $N$, $A$ is the amplitude and $B$ is a constant. Figure \ref{fig:meanof10simu3sig} also shows the corresponding $3\sigma$ error region. The error bars on the data points are the standard deviations of the values of $ncp\_prior$ across all simulations.

For comparison, we also plot 2 other calibrations used in \citet{scargle2013studies}. They are derived via similar simulations, but for Gaussian noise, and only consider $N$ values up to 1024. We show the $3\sigma$ error region to accommodate the noise and to minimize false positives (recall that higher values of $ncp\_prior$ imply that the algorithm is less susceptible to false positives). Roughly, this means that we are $3\sigma$ confident that our calibration yields $p_0 \leq 0.05$. Our final calibration is the upper bound of the $3\sigma$ fit.

\label{sec:App_A}

\begin{table}[htp]
\centering
\renewcommand{\arraystretch}{0.75}
\begin{tabular}{ccccccccc}
\hline
ObsID & Exp    & Quie. count rate & Fstart    & Fend      & Dur   & Mean count rate & Emitted energy \\
      & (ks)   &($\times 10^{-3}$ ct s$^{\text{-1}}$)& (MJD)     & (MJD)     & (s)   & (ct s$^{\text{-1}}$)          & (2-10 keV, $10^{37}$ erg) \\
\hline
13850 & 60.0  &  $5.9 \pm 0.3$   &           &           &       &                 &         \\
14392 & 59.0  &  $6.9 \pm 0.5$   & 55966.433 & 55966.458 & 2109  &$0.021 \pm 0.003$&  3.74       \\
      &        &                  & 55966.602 & 55966.667 & 5641  &$0.145 \pm 0.004$&  101.1       \\
14394 & 17.4  &  $6.8 \pm 0.6$   &           &           &       &                 &         \\
14393 & 41.1  &  $8.0 \pm 0.4$   &           &           &       &                 &         \\
13856 & 39.7  &  $5.6 \pm 0.4$   &           &           &       &                 &         \\
13857 & 39.5  &  $6.7 \pm 0.5$   &           &           &       &                 &         \\
13854 & 22.7  &  $8 \pm 1    $   & 56006.487 & 56006.494 & 669   &$0.054 \pm 0.009$&  3.97       \\
      &        &                  & 56006.528 & 56006.538 & 817   &$0.055 \pm 0.008$&  4.96       \\
      &        &                  & 56006.586 & 56006.596 & 944   &$0.053 \pm 0.007$&  5.45       \\
      &        &                  & 56006.682 & 56006.687 & 402   &$0.11   \pm 0.02$&  5.08       \\
14413 & 14.7  &  $6.5 \pm 0.7$   &           &           &       &                 &         \\
13855 & 19.9  &  $6.9 \pm 0.6$   &           &           &       &                 &         \\
14414 & 20.0  &  $5.9 \pm 0.5$   &           &           &       &                 &         \\
13847 & 153.4 &  $6.3 \pm 0.3$   & 56048.511 & 56048.549 & 3251  &$0.024 \pm 0.003$&   7.56      \\
14427 & 79.9  &  $5.7 \pm 0.5$   & 56054.097 & 56054.151 & 4679  &$0.020 \pm 0.002$&   8.71      \\
      &        &                  & 56054.468 & 56054.490 & 1899  &$0.020 \pm 0.003$&   3.46     \\
13848 & 97.8  &  $6.4 \pm 0.3$   &           &           &       &                 &         \\
13849 & 178.3 &  $6.8 \pm 0.3$   & 56058.690 & 56058.704 & 1172  &$0.028 \pm 0.005$&   3.19      \\
      &        &                  & 56059.008 & 56059.064 & 4891  &$0.015 \pm 0.002$&   5.36      \\
      &        &                  & 56059.315 & 56059.330 & 1244  &$0.025 \pm 0.004$&   2.96      \\
      &        &                  & 56060.137 & 56060.166 & 2435  &$0.055 \pm 0.005$&   15.46      \\
13846 & 55.8  &  $6.1 \pm 0.3$   &           &           &       &                 &         \\
14438 & 25.3  &  $6.3 \pm 0.5$   &           &           &       &                 &         \\
13845 & 133.3 &  $5.9 \pm 0.2$   & 56066.576 & 56066.595 & 1633  &$0.018 \pm 0.003$&   2.53      \\
      &        &                  & 56067.864 & 56068.026 & 14006 &$0.017 \pm 0.001$&   20.38     \\
14460 & 23.8  &  $5.6 \pm 0.7$   &           &           &       &                 &         \\
13844 & 19.8  &  $5.9 \pm 0.5$   &           &           &       &                 &         \\
14461 & 50.6  &  $7.0 \pm 0.4$   &           &           &       &                 &         \\
13853 & 73.3  &  $5.7 \pm 0.3$   &           &           &       &                 &         \\
13841 & 44.8  &  $6.2 \pm 0.4$   &           &           &       &                 &         \\
14465 & 44.2  &  $5.8 \pm 0.5$   & 56126.977 & 56127.034 & 4911$^{\text{a}}$  &$0.019 \pm 0.002$&  8.25       \\
      &        &                  & 56127.177 & 56127.203 & 2255  &$0.018 \pm 0.003$&  3.56       \\
14466 & 45.0  &  $7.0 \pm 0.4$   & 56128.552 & 56128.555 & 286$^{\text{a}}$   &$0.08 \pm 0.02$&     2.62    \\
13842 & 191.6 &  $6.1 \pm 0.3$   & 56130.187 & 56130.225 & 3246  &$0.039 \pm 0.003$&  13.84       \\
      &        &                  & 56130.908 & 56130.919 & 925   &$0.055 \pm 0.008$&  5.87       \\
      &        &                  & 56131.494 & 56131.585 & 7865  &$0.021 \pm 0.002$&  15.63       \\
13839 & 175.5 &  $6.5 \pm 0.2$   & 56132.389 & 56132.399 & 856   &$0.051 \pm 0.008$&  4.95       \\
      &        &                  & 56133.999 & 56134.180 & 15614 &$0.026 \pm 0.001$&  40.33       \\
13840 & 162.2 &  $6.7 \pm 0.2$   & 56136.457 & 56136.507 & 4271  &$0.015 \pm 0.002$&  4.42       \\
      &        &                  & 56136.627 & 56136.655 & 2346  &$0.019 \pm 0.003$&  3.72       \\
14432 & 74.2  &  $5.9 \pm 0.3$   & 56138.558 & 56138.609 & 4482$^{\text{a}}$  &$0.013 \pm 0.002$&  3.88       \\
      &        &                  & 56139.374 & 56139.416 & 3651$^{\text{a}}$  &$0.053 \pm 0.004$&  22.57      \\
13838 & 99.1  &  $6.5 \pm 0.3$   & 56141.013 & 56141.049 & 3067  &$0.055 \pm 0.004$&  19.48       \\
13852 & 155.9 &  $6.9 \pm 0.3$   & 56143.317 & 56143.331 & 1215  &$0.053 \pm 0.007$&  7.22       \\
      &        &                  & 56144.330 & 56144.350 & 1724  &$0.022 \pm 0.004$&  3.45       \\
14439 & 111.3 &  $6.3 \pm 0.2$   & 56147.132 & 56147.147 & 1338  &$0.026 \pm 0.004$&  3.51       \\
14462 & 133.8 &  $5.9 \pm 0.3$   & 56207.180 & 56207.191 & 947   &$0.031 \pm 0.006$&  3.07       \\
      &        &                  & 56208.187 & 56208.220 & 2822  &$0.025 \pm 0.003$&  7.02       \\
14463 & 30.4  &  $6.7 \pm 0.6$   & 56216.240 & 56216.247 & 525   &$0.12  \pm 0.02$ &  7.89       \\
13851 & 106.8 &  $5.3 \pm 0.3$   & 56217.096 & 56217.097 & 152   &$0.07 \pm 0.02$  &  1.34       \\
      &        &                  & 56217.814 & 56217.878 & 5486  &$0.074 \pm 0.004$&  51.76      \\
15568 & 35.9  &  $6.3 \pm 0.4$   &           &           &       &                 &         \\
13843 & 120.4 &  $6.4 \pm 0.3$   & 56223.381 & 56223.476 & 8174  &$0.034 \pm 0.002$&  29.02       \\
15570 & 68.4  &  $6.0 \pm 0.3$   & 56225.233 & 56225.262 & 2436  &$0.032 \pm 0.004$&  8.10       \\
14468 & 145.9 &  $5.8 \pm 0.2$   & 56230.298 & 56230.338 & 3504  &$0.024 \pm 0.003$&  8.38       \\
      &        &                  & 56231.567 & 56231.592 & 2178  &$0.030 \pm 0.004$&  6.90       \\
\hline
\end{tabular}
\caption{List of all $Chandra$ XVP observations. Listed for each observation are the exposure (taken as the total time spanned by the Bayesian Blocks), the flare start and end times and the duration. The flare mean count rates are pile-up corrected but not quiescence subtracted and we indicate their Poisson error bars. The energies are from 2-10 keV (not 2-8 keV like the mean count rates), are pile-up corrected and quiescence subtracted. \\\hspace{\textwidth} $^{\text{a}}$ Flare truncated by the beginning or end of the observation. }
\label{tab:XVP-flares}
\end{table}

\begin{table}[htp]
\centering
\begin{tabular}{ccccccc}
\hline
ObsID & Exposure & Fstart    & Fend      & Dura & Mean cr & Emitted energy \\
      & (ks)       & (MJD)       & (MJD)       & (s)    & (ct s$^{\text{-1}}$)    &  (2-10 keV, $10^{37}$ erg) \\
\hline
15041 & 50.0    & $56500.146$ & $56500.158$ & $1041$ & $0.035 \pm 0.006$ & 2.03\\
      &          &$56500.460$ & $56500.470$ & $850 $ & $0.033 \pm 0.006$ & 1.55\\
15042 & 49.2    & $56516.357$ & $56516.371$ & $1262$ & $0.028 \pm 0.005$ & 1.84\\
      &          &$56516.405$ & $56516.441$ & $3109$ & $0.028 \pm 0.003$ & 4.60\\
15043 & 50.0    & $56549.085$ & $56549.108^{\text{a}}$ & $2036$ & $0.75  \pm 0.02 $ & 127.86\\
      &          & $56549.108^{\text{a}}$ & $56549.149$ & $3503$ & $0.49  \pm 0.01 $ & 142.91\\
15045 & 50.0    & $56593.675$ & $56593.702$ & $2300$ & $0.033 \pm 0.004$ & 4.31\\
      &          &$56593.831$ & $56593.842$ & $946 $ & $0.031 \pm 0.006$ & 1.63\\
16508 & 47.8    & $56710.024$ & $56710.048$ & $2066^{\text{b}}$ & $0.027\pm 0.004$ & 3.02\\
16217 & 37.8    & $56899.488$ & $56899.546$ & $4979$ & $0.019\pm 0.002$ & 5.07\\
16218 & 40.0    & $56950.557$ & $56950.606^{\text{a}}$ & $4189$ & $0.191 \pm 0.007$ & 58.79\\
      &          & $56950.606^{\text{a}}$ & $56950.630$ & $2055$ & $0.031 \pm 0.004$ & 3.81\\
16963 & 24.8    & $57066.252$ & $57066.262$ & $869 $ & $0.044 \pm 0.007$ & 2.51\\
16966 & 24.6    & $57156.501$ & $57156.537$ & $3131$ & $0.054 \pm 0.004$ & 11.14\\
18731 & 86.3    & $57581.947$ & $57581.957$ & $904 $ & $0.036 \pm 0.006$ & 2.06\\
18732 & 84.4    & $57587.623$ & $57587.653$ & $2564$ & $0.022 \pm 0.003$ & 3.36\\
20041 & 33.9    & $57854.352$ & $57854.379$ & $2287$ & $0.103 \pm 0.007$ & 16.70\\
19703 & 89.0    & $57949.959$ & $57950.009$ & $4610^{\text{b}}$ & $0.015 \pm 0.002$ & 3.40\\
      &          & $57950.547$ & $57950.561$ & $1210$ & $0.033 \pm 0.005$ & 2.66\\
20346 & 33.0    & $58232.206$ & $58232.219^{\text{a}}$ & $1082$ & $0.12 \pm 0.01 $ & 9.39\\
      &          & $58232.219^{\text{a}}$ & $58232.246$ & $2373$ & $0.117 \pm 0.007$ & 20.48\\
\hline
\end{tabular}
\caption{List of all $Chandra$ flares from our kept Post-XVP flare dataset. Listed for each flare are the observation start date, the exposure (taken as the total time spanned by the Bayesian Blocks), the flare start and end times and the duration. The flare mean count rates are pile-up corrected but not quiescence-subtracted and we indicate their Poisson error bars. The energies are from 2-10 keV (not 2-8 keV like the mean count rates), are pile-up corrected and quiescence-subtracted. They are obtained from the spectral model used by \cite{neilsen2013chandra} via a method explained in Section \ref{sec:crtoenergies}. \\\hspace{\textwidth} $^{\text{a}}$ Flares showing important substructure, which could indicate the presence of two flare if we use the criterion explained in Section \ref{sec:BB_class}. \\\hspace{\textwidth} $^{\text{b}}$ Flare truncated by the beginning or end of the observation. }
\label{tab:Post-XVP-flares}
\end{table}

\section{Comparison with other works}
\label{sec:compar}
In this section, we compare our flares with \citet{neilsen2013chandra}, \citet{ponti2015fifteen} and \citet{mossoux2017sixteen}. To compare energies with \citet{ponti2015fifteen}, we convert from absorbed to unabsorbed values using the scaling between the unabsorbed fluences and absorbed fluxes of Table 1 of \citet{neilsen2013chandra} assuming a distance of 8 kpc. This should be sufficient for comparison purposes. For the comparisons in this section, we use our flares as obtained without the flare splitting criterion (see Section \ref{sec:BB_class}). Table \ref{tab:Compare} shows every flare detected or missed by every work considered and Figures \ref{fig:DurComp} and \ref{fig:FluComp} compare the different flares' durations and energies of each author. From Figure \ref{fig:FluComp}, our work (green '+') shows remarkable agreement with \citet{neilsen2013chandra} (red triangles) for the XVP dataset. This is expected since we also scale our flares' fluxes to the brightest flare of 2012 from \citet{nowak2012chandra}. The work of \citet{mossoux2017sixteen} is represented by the blue 'x' and the flares of \citet{ponti2015fifteen} are shown as black dots. 

\subsection{XVP flares}

\begin{table}[htp]
\centering
\renewcommand{\arraystretch}{0.71}
\begin{tabular}{cccccc}
\hline
ObsID & Flare index & This work  & N13        & P15                           & M17                           \\
\hline
14392 & 1           & \checkmark & \checkmark &                               & \checkmark                    \\
      & 2           & \checkmark & \checkmark & \checkmark                    & \checkmark                    \\
13857 & 3           &            &            &                               & \checkmark                    \\
13854 & 4           & \checkmark & \checkmark & \checkmark                    & \checkmark                    \\
      & 5           & \checkmark & \checkmark & \checkmark                    & \checkmark                    \\
      & 6           & \checkmark & \checkmark & \checkmark                    & \checkmark                    \\
      & 7           & \checkmark & \checkmark & \checkmark                    & \checkmark                    \\
13847 & 8           & \checkmark & \checkmark & \checkmark                    & \checkmark                    \\
      & 9           &            & \checkmark &                               &                               \\
14427 & 10          & \checkmark & \checkmark & \checkmark                    & \checkmark                    \\
      & 11          & \checkmark &            & \checkmark                    & \checkmark                    \\
13849 & 12          & \checkmark & \checkmark & \checkmark                    & \checkmark                    \\
      & 13          & \checkmark & \checkmark & \checkmark                    & \checkmark                    \\
      & 14          & \checkmark & \checkmark &                               & \checkmark                    \\
      & 15          & \checkmark & \checkmark & \checkmark                    & \checkmark                    \\
13845 & 16          & \checkmark &            &                               & \checkmark                    \\
      & 17          & \checkmark & \checkmark & \checkmark                    & \checkmark                    \\
13853 & 18          &            & \checkmark &                               &                               \\
14465 & 19          & \checkmark & \checkmark & \checkmark                    & \checkmark                    \\
      & 20          & \checkmark & \checkmark & \checkmark                    & \checkmark                    \\
14466 & 21          & \checkmark & \checkmark & \checkmark                    & \checkmark                    \\
      & 22          &            &            &                               & \checkmark                    \\
13842 & 23          & \checkmark & \checkmark & \checkmark                    & \checkmark                    \\
      & 24          & \checkmark & \checkmark & \checkmark                    & \checkmark                    \\
      & 25          & \checkmark & \checkmark & \checkmark                    & \checkmark                    \\
13839 & 26          & \checkmark & \checkmark & \checkmark                    & \checkmark                    \\
      & 27          &            & \checkmark &                               & \checkmark                    \\
      & 28          & \checkmark & \checkmark & \checkmark                    & \checkmark                    \\
13840 & 29          & \checkmark &            & \checkmark                    & \checkmark                    \\
      & 30          & \checkmark &            &                               & \checkmark                    \\
14432 & 31          & \checkmark &            & \checkmark                    & \checkmark                    \\
      & 32          & \checkmark & \checkmark & \checkmark                    & \checkmark                    \\
13838 & 33          & \checkmark & \checkmark & \checkmark                    & \checkmark                    \\
13852 & 34          & \checkmark & \checkmark & \checkmark                    & \checkmark                    \\
      & 35          &            &            & \checkmark                    &                               \\
      & 36          & \checkmark & \checkmark & \checkmark                    & \checkmark                    \\
14439 & 37          & \checkmark & \checkmark & \checkmark                    & \checkmark                    \\
14462 & 38          & \checkmark & \checkmark & \checkmark                    & \checkmark                    \\
      & 39          & \checkmark & \checkmark & \checkmark                    & \checkmark                    \\
14463 & 40          & \checkmark & \checkmark & \checkmark                    & \checkmark                    \\
13851 & 41          & \checkmark & \checkmark &                               & \checkmark                    \\
      & 42          & \checkmark & \checkmark & \checkmark                    & \checkmark                    \\
15568 & 43          &            &            & \checkmark                    & \checkmark                    \\
13843 & 44          & \checkmark & \checkmark & \checkmark                    & \checkmark                    \\
15570 & 45          & \checkmark & \checkmark & \checkmark                    & \checkmark                    \\
14468 & 46          & \checkmark & \checkmark & \checkmark                    & \checkmark                    \\
      & 47          &            & \checkmark &                               &                               \\
      & 48          & \checkmark & \checkmark & \checkmark                    & \checkmark                    \\
\hline
\hline
15041 & 49          & \checkmark & --         & \checkmark                    & \checkmark                    \\
      & 50          & \checkmark & --         &                               & \checkmark                    \\
15042 & 51          & \checkmark & --         & \checkmark $^{\text{a}}$      & \checkmark                    \\
      & 52          & \checkmark & --         & $^{\text{a}}$                 & \checkmark                    \\
14945 & 53          &            & --         & \checkmark                    &                               \\
15043 & 54          & \checkmark & --         & \checkmark                    & \checkmark                    \\
14943 & 55          &            & --         & \checkmark                    &                               \\
15045 & 56          & \checkmark & --         & \checkmark                    & \checkmark                    \\
      & 57          & \checkmark & --         & \checkmark                    & \checkmark                    \\
16508 & 58          & \checkmark & --         & \checkmark                    &                               \\
16217 & 59          & \checkmark & --         & \checkmark                    &                               \\
16218 & 60          & \checkmark$^{\text{a}}$ & --         & \checkmark                    & \checkmark $^{\text{a}}$      \\
      & 61          & $^{\text{a}}$           & --         & \checkmark                    & $^{\text{a}}$                 \\
16963 & 62          & \checkmark & --         & --                            & \checkmark                    \\
16966 & 63          & \checkmark & --         & --                            & \checkmark  \\ 
\hline
\end{tabular}
\caption{List of all the flares reported in this work, \citet{neilsen2013chandra} (N13), \citet{ponti2015fifteen} (P15) and \citet{mossoux2017sixteen} (M17) within the obervations considered in this analysis. We show our results without the flare splitting criterion. We stop at ObsID 16966 since more recent ObsIDs were not available at the time of publishing the other papers. \\\hspace{\textwidth} $^{\text{a}}$ These authors report these two flares as a single one.}
\label{tab:Compare}
\end{table}

\citet{neilsen2013chandra} report 39 XVP flares and we find 40. 35 flares are in common. We find 5 long (durations above 1600 s) and faint (0.019 ct s$^{\text{-1}}$ and less) flares that they missed in ObsIDs 14427, 13845, 13840 (2 flares in this ObsID) and 14432 (flares \# 11, 16, 29, 30 and 31). On the other hand, they report 4 short (durations from 500 s to 1200 s) and faint (between 8 and 15 counts, or 1.15 and 2 $\times 10^{37}$ erg) flares that we do not find in ObsIDs 13847, 13853, 13839 and 14468 (flares \# 9, 18, 27 and 47). Our results agree for the most significant flares, but some differences exist for the weaker flares, which can be explained by the different detection methods used (long and faint flares are more easily detected by Bayesian Blocks, but short and faint flares are best found by direct Gaussian fitting). 

\cite{ponti2015fifteen} find 37 XVP flares, of which 35 are in common. We find 4 faint (unabsorbed fluences below $3.7 \times 10^{37}$ erg) flares that they miss in ObsIDs 14392 (\# 1), 13849 (\# 14), 13845 (\# 16) and 13840 (\# 30). They are most likely missed because these authors limit themselves to the 0th order. They also report 2 flares that we do not detect in ObsIDs 13852 (\# 35, with a count rate of 0.0049 ct s$^{\text{-1}}$, consistent with quiescence and not found by \citet{mossoux2017sixteen} nor \citet{neilsen2013chandra}) and 15568 (\# 43). The flare at ObsID 15568 is also found by \citet{mossoux2017sixteen} but not by us nor \citet{neilsen2013chandra}. It is rather faint ($\sim$ 2x quiescence) and its detection is thus sensitive to what different authors consider to be significant.

\citet{mossoux2017sixteen} find 44 XVP flares, of which we have 40 in common. The 4 flares that we do not find are in ObsIDs 13857, 14466, 13839, and 15568 (flares \# 3, 22, 27 and 43). Even though we find blocks in ObsIDs 13857, 14466, and 15568, they are not considered flares because of our significance criterion. The reported flares at ObsIDs 13857 and 14466 are short (durations of 471 s and 380 s) and the one at ObsID 15568 lasts >4907 s but is very faint (0.006 ct s$^{\text{-1}}$ above quiescence). This translates to rather high error bars, which explains why our method doesn't consider them flares. The flare at ObsID 13839 is undetected by our Bayesian Blocks and also by \citet{ponti2015fifteen}, but it is on the short end (750 s according to \citet{neilsen2013chandra} and 1411 s according to \citet{mossoux2017sixteen}).

\subsection{Post-XVP flares}
\begin{figure}[h!]
    \centering
    \includegraphics[width=0.9\textwidth]{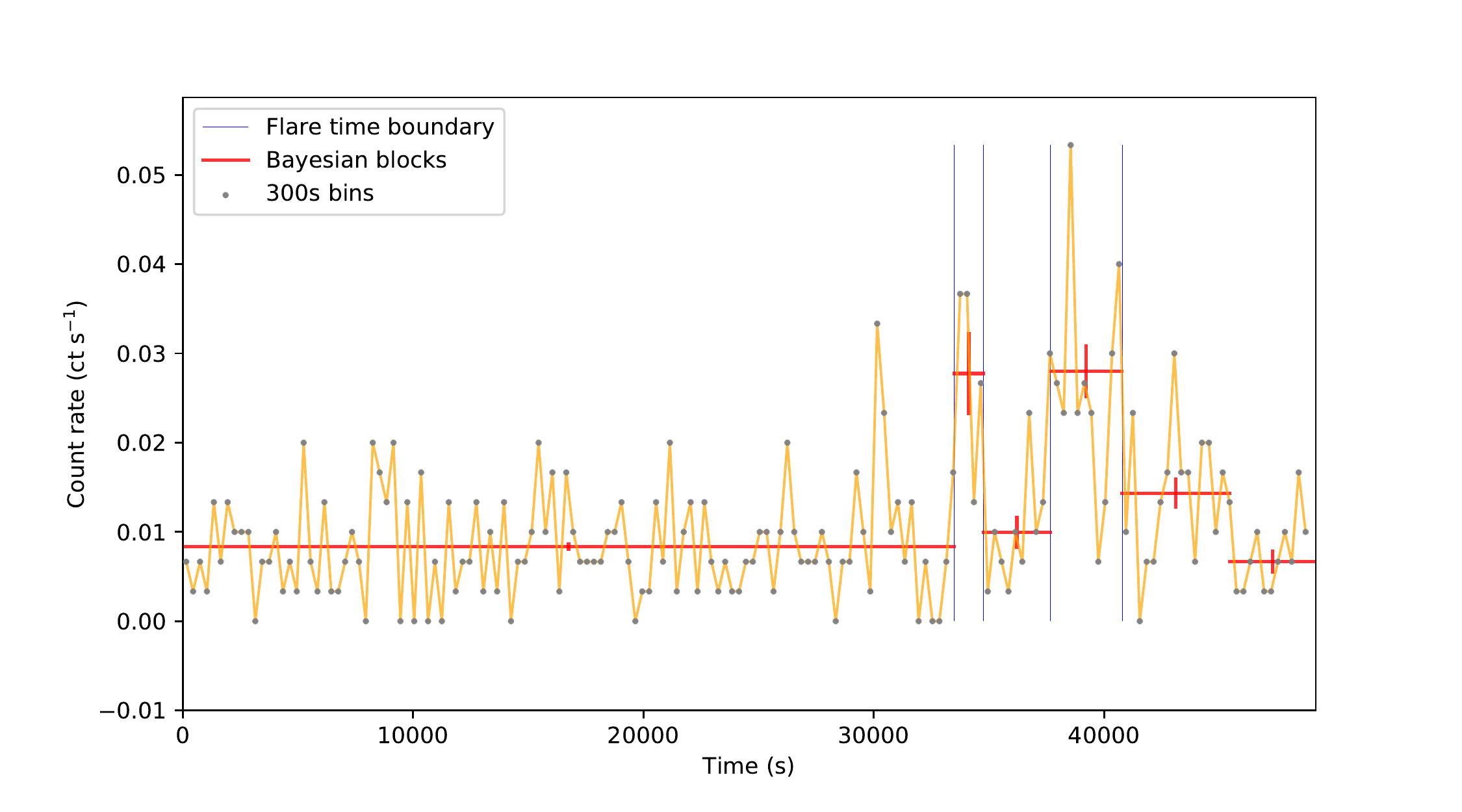}
    \caption{Light curve of ObsID 15042. Notice the block separating the two flares.}
    \label{fig:15042}
\end{figure}

\begin{figure}[h!]
    \centering
    \includegraphics[width=0.9\textwidth]{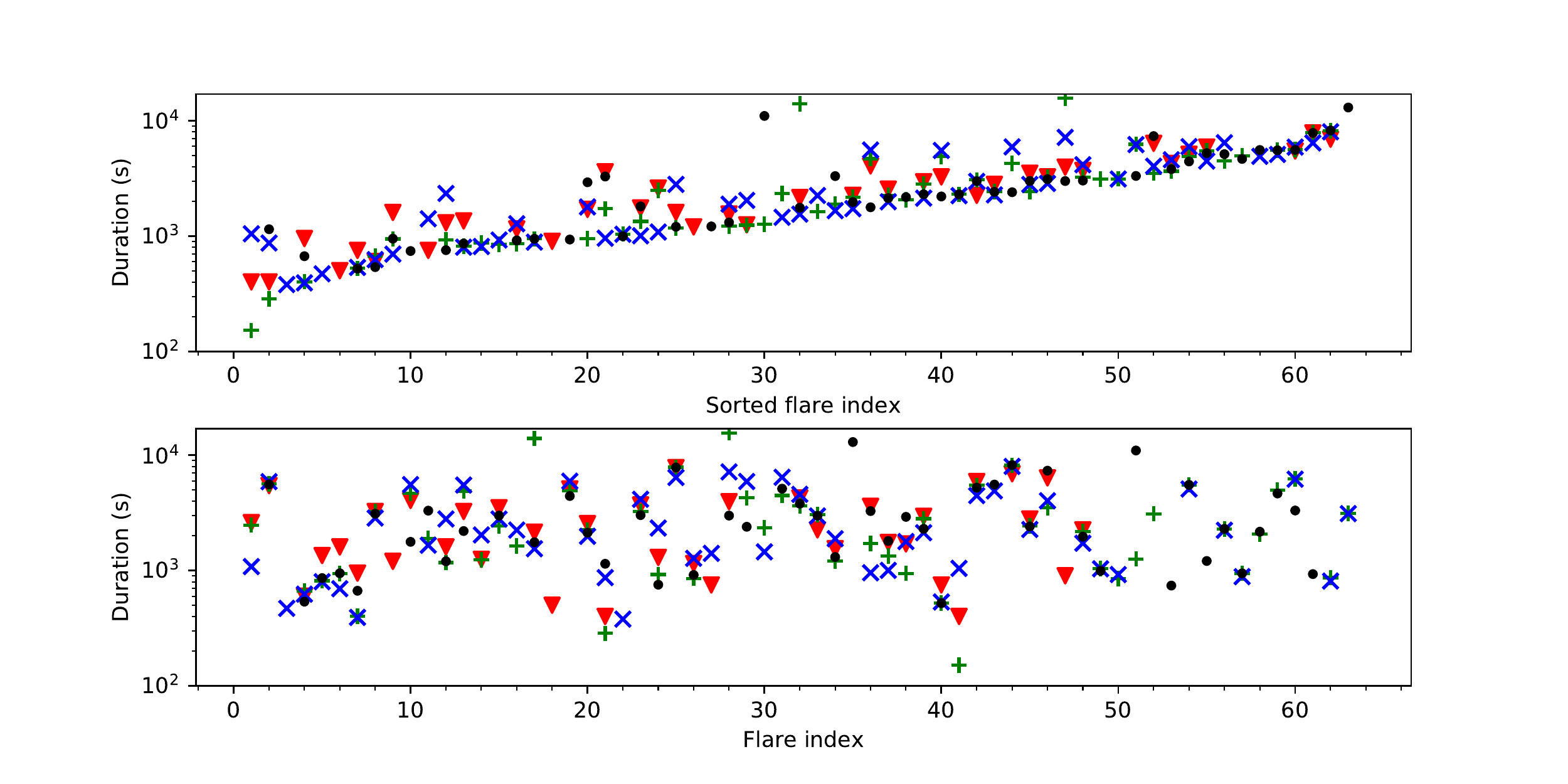}
    \caption{Durations of the flares detected in this work (green '+'), \citet{neilsen2013chandra} (red triangles), \citet{mossoux2017sixteen} (blue 'x') and \citet{ponti2015fifteen} (black dots). $Top$: Flares ordered by the shortest duration seen for each flare. $Bottom:$ Flares ordered by the indices of Table \ref{tab:Compare}.}
    \label{fig:DurComp}
\end{figure}

\begin{figure}[h!]
    \centering
    \includegraphics[width=0.9\textwidth]{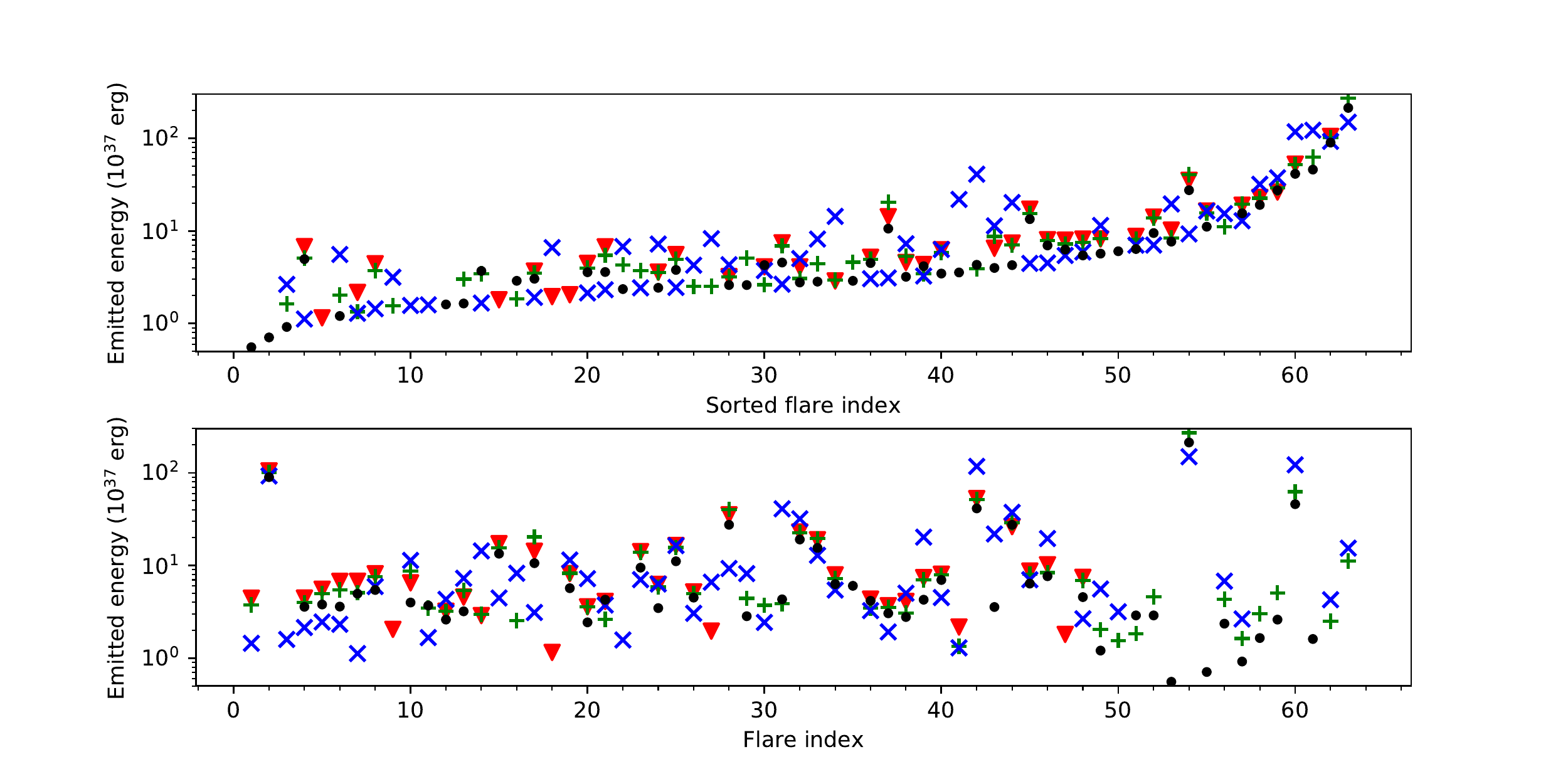}
    \caption{Like Figure \ref{fig:DurComp} but for flare energies. For \citet{ponti2015fifteen}, we converted from absorbed to unabsorbed values using the scaling between the unabsorbed fluences and absorbed fluxes of Table 1 of \citet{neilsen2013chandra} assuming a distance of 8 kpc. This should be sufficient for comparison purposes.}
    \label{fig:FluComp}
\end{figure}

\citet{ponti2015fifteen} report 10 flares for the Post-XVP observations that we share. There are substantial differences with what we find. For ObsID 15041, we and \citet{mossoux2017sixteen} find two flares (flares \# 49 and 50), but \citet{ponti2015fifteen} only find the first one, which is faint (emitted energy of $1.55 \times 10^{37}$ erg). Likewise, we find two flares in ObsID 15042 (flares \# 51 and 52), but \citet{ponti2015fifteen} report them as one and \citet{mossoux2017sixteen} do not find any. We argue that there are two distinct flares since we can see in Figure \ref{fig:15042} that the two flares are separated by a non-flaring block. \citet{ponti2015fifteen} also report 2 faint flares in ObsIDs 14945 and 14943 (flares \# 53 and 55), but \citet{mossoux2017sixteen} and us do not find them. \citet{ponti2015fifteen} also report a flare in ObsID 16508 (\# 58) and another one in ObsID 16217 (\# 59) that \citet{mossoux2017sixteen} do not find, but we do. Briefly, our results are a combination of these authors and differences occur based on the criterion that selects what is considered a flaring block and how the Bayesian Blocks are calibrated.

\section{Pile-up treatment in simulated gratings light curves}
\label{sec:App_puGrating}
To simulate a flare with a given mean unpiled count rate in the HETG/ACIS-S/0+1st instrument mode, we need the unpiled 0th/1st order flare and quiescence ratios as pile-up occurs on the observed total 0th order count rate only. We assume these ratios to be the median of the ones observed in the XVP dataset, $i.e.$, $\sim 1.6$ and $0.45$ for flare and quiescence, respectively.

The unpiling process requires more attention since the (piled) 0th/1st order flare ratio changes with flare count rate due to pile-up. This phenomenon can be taken into account via a calibration that computes the piled 0th/1st order flare ratio for many different unpiled count rates.\footnote{Assuming a quiescent count rate of 0.0063 ct s$^{\text{-1}}$ (the median value of all XVP HETG observations) and a quiescent 0th/1st order ratio of 0.45.} When the Bayesian Blocks algorithm detects a flare mean count rate, we use that calibration to find the corresponding piled 0th/1st order flare ratio. This information is used to infer the total unpiled flare count rate.

\clearpage
\bibliographystyle{aasjournal}
\bibliography{main.bib}

\begin{thebibliography}{}
\expandafter\ifx\csname natexlab\endcsname\relax\def\natexlab#1{#1}\fi

\bibitem[{Arnaud(1996)}]{arnaud1996xspec}
Arnaud, K. 1996, in Astronomical Data Analysis Software and Systems V, Vol.
  101, 17

\bibitem[{Baganoff {et~al.}(2001)Baganoff, Bautz, Brandt, Chartas, Feigelson,
  Garmire, Maeda, Morris, Ricker, Townsley, {et~al.}}]{baganoff2001rapid}
Baganoff, F., Bautz, M., Brandt, W.~N., {et~al.} 2001, Nature, 413, 45

\bibitem[{Baganoff {et~al.}(2003)Baganoff, Maeda, Morris, Bautz, Brandt, Cui,
  Doty, Feigelson, Garmire, Pravdo, {et~al.}}]{baganoff2003chandra}
Baganoff, F.~K., Maeda, Y., Morris, M., {et~al.} 2003, The Astrophysical
  Journal, 591, 891

\bibitem[{Ball {et~al.}(2016)Ball, {\"O}zel, Psaltis, \&
  Chan}]{ball2016particle}
Ball, D., {\"O}zel, F., Psaltis, D., \& Chan, C.-k. 2016, The Astrophysical
  Journal, 826, 77

\bibitem[{Ball {et~al.}(2018)Ball, {\"O}zel, Psaltis, Chan, \&
  Sironi}]{ball2018properties}
Ball, D., {\"O}zel, F., Psaltis, D., Chan, C.-K., \& Sironi, L. 2018, The
  Astrophysical Journal, 853, 184

\bibitem[{Barriere {et~al.}(2014)Barriere, Tomsick, Baganoff, Boggs,
  Christensen, Craig, Dexter, Grefenstette, Hailey, Harrison,
  {et~al.}}]{barriere2014nustar}
Barriere, N.~M., Tomsick, J.~A., Baganoff, F.~K., {et~al.} 2014, The
  Astrophysical Journal, 786, 46

\bibitem[{Belanger {et~al.}(2005)Belanger, Goldwurm, Melia, Ferrando, Grosso,
  Porquet, Warwick, \& Yusef-Zadeh}]{belanger2005repeated}
Belanger, G., Goldwurm, A., Melia, F., {et~al.} 2005, The Astrophysical
  Journal, 635, 1095

\bibitem[{Boyce {et~al.}(2019)Boyce, Haggard, Witzel, Willner, Neilsen, Hora,
  Markoff, Ponti, Baganoff, Becklin, {et~al.}}]{boyce2019simultaneous}
Boyce, H., Haggard, D., Witzel, G., {et~al.} 2019, The Astrophysical Journal,
  871, 161

\bibitem[{{\v{C}}ade{\v{z}} {et~al.}(2008){\v{C}}ade{\v{z}}, Calvani, \&
  Kosti{\'c}}]{vcadevz2008tidal}
{\v{C}}ade{\v{z}}, A., Calvani, M., \& Kosti{\'c}, U. 2008, Astronomy \&
  Astrophysics, 487, 527

\bibitem[{{Coti Zelati} {et~al.}(2017){Coti Zelati}, {Rea}, {Turolla}, {Pons},
  {Papitto}, {Esposito}, {Israel}, {Campana}, {Zane}, {Tiengo}, {Mignani},
  {Mereghetti}, {Baganoff}, {Haggard}, {Ponti}, {Torres}, {Borghese}, \&
  {Elfritz}}]{2017MNRAS.471.1819C}
{Coti Zelati}, F., {Rea}, N., {Turolla}, R., {et~al.} 2017, MNRAS, 471, 1819

\bibitem[{Davis(1984)}]{davis1984piecewise}
Davis, M.~H. 1984, Journal of the Royal Statistical Society: Series B
  (Methodological), 46, 353

\bibitem[{Degenaar {et~al.}(2013)Degenaar, Miller, Kennea, Gehrels, Reynolds,
  \& Wijnands}]{degenaar2013x}
Degenaar, N., Miller, J., Kennea, J., {et~al.} 2013, The Astrophysical Journal,
  769, 155

\bibitem[{Dibi {et~al.}(2014)Dibi, Markoff, Belmont, Malzac, Barri{\`e}re, \&
  Tomsick}]{dibi2014exploring}
Dibi, S., Markoff, S., Belmont, R., {et~al.} 2014, Monthly Notices of the Royal
  Astronomical Society, 441, 1005

\bibitem[{Dibi {et~al.}(2016)Dibi, Markoff, Belmont, Malzac, Neilsen, \&
  Witzel}]{dibi2016using}
---. 2016, Monthly Notices of the Royal Astronomical Society, 461, 552

\bibitem[{Do {et~al.}(2019)Do, Witzel, Gautam, Chen, Ghez, Morris, Becklin,
  Ciurlo, Hosek~Jr, Martinez, {et~al.}}]{do2019unprecedented}
Do, T., Witzel, G., Gautam, A.~K., {et~al.} 2019, arXiv preprint
  arXiv:1908.01777

\bibitem[{Dodds-Eden {et~al.}(2009)Dodds-Eden, Porquet, Trap, Quataert,
  Haubois, Gillessen, Grosso, Pantin, Falcke, Rouan,
  {et~al.}}]{dodds2009evidence}
Dodds-Eden, K., Porquet, D., Trap, G., {et~al.} 2009, The Astrophysical
  Journal, 698, 676

\bibitem[{Eckart {et~al.}(2009)Eckart, Baganoff, Morris, Kunneriath,
  Zamaninasab, Witzel, Sch{\"o}del, Garc{\'\i}a-Mar{\'\i}n, Meyer, Bower,
  {et~al.}}]{eckart2009modeling}
Eckart, A., Baganoff, F., Morris, M., {et~al.} 2009, Astronomy \& Astrophysics,
  500, 935

\bibitem[{EHT~Collaboration {et~al.}(2019)EHT~Collaboration, Alberdi, Alef,
  Asada, Azulay, Baczko, Ball, Baloković, Barrett, Bintley, Blackburn,
  {et~al.}}]{EHTm87-I}
EHT~Collaboration, A.~K., Alberdi, A., Alef, W., {et~al.} 2019, The
  Astrophysical Journal, 875, L1

\bibitem[{Genzel {et~al.}(2010)Genzel, Eisenhauer, \&
  Gillessen}]{genzel2010galactic}
Genzel, R., Eisenhauer, F., \& Gillessen, S. 2010, Reviews of Modern Physics,
  82, 3121

\bibitem[{Gillessen {et~al.}(2012)Gillessen, Genzel, Fritz, Quataert, Alig,
  Burkert, Cuadra, Eisenhauer, Pfuhl, Dodds-Eden, {et~al.}}]{gillessen2012gas}
Gillessen, S., Genzel, R., Fritz, T., {et~al.} 2012, Nature, 481, 51

\bibitem[{Gillessen {et~al.}(2013{\natexlab{a}})Gillessen, Genzel, Fritz,
  Eisenhauer, Pfuhl, Ott, Cuadra, Schartmann, \& Burkert}]{gillessen2013new}
---. 2013{\natexlab{a}}, The Astrophysical Journal, 763, 78

\bibitem[{Gillessen {et~al.}(2013{\natexlab{b}})Gillessen, Genzel, Fritz,
  Eisenhauer, Pfuhl, Ott, Schartmann, Ballone, \&
  Burkert}]{gillessen2013pericenter}
Gillessen, S., Genzel, R., Fritz, T.~K., {et~al.} 2013{\natexlab{b}}, The
  Astrophysical Journal, 774, 44

\bibitem[{Gillessen {et~al.}(2017)Gillessen, Plewa, Eisenhauer, Sari, Waisberg,
  Habibi, Pfuhl, George, Dexter, von Fellenberg,
  {et~al.}}]{gillessen2017update}
Gillessen, S., Plewa, P., Eisenhauer, F., {et~al.} 2017, The Astrophysical
  Journal, 837, 30

\bibitem[{Gillessen {et~al.}(2018)Gillessen, Plewa, Widmann, von Fellenberg,
  Schartmann, Habibi, Rosales, Baub{\"o}ck, Dexter, Gao,
  {et~al.}}]{gillessen2018detection}
Gillessen, S., Plewa, P., Widmann, F., {et~al.} 2018, arXiv preprint
  arXiv:1812.01416

\bibitem[{Goldwurm {et~al.}(2003)Goldwurm, Brion, Goldoni, Ferrando, Daigne,
  Decourchelle, Warwick, \& Predehl}]{goldwurm2003new}
Goldwurm, A., Brion, E., Goldoni, P., {et~al.} 2003, The Astrophysical Journal,
  584, 751

\bibitem[{Gravity~Collaboration {et~al.}(2018)Gravity~Collaboration, Amorim,
  Baub{\"o}ck, Berger, Bonnet, Brandner, Cl{\'e}net, du~Foresto, de~Zeeuw,
  Deen, {et~al.}}]{abuter2018detection}
Gravity~Collaboration, A.~R., Amorim, A., Baub{\"o}ck, M., {et~al.} 2018,
  Astronomy \& Astrophysics, 618, L10

\bibitem[{Haggard {et~al.}(2019)}]{Haggard19}
Haggard, D., {et~al.} 2019, in preparation

\bibitem[{Kawashima {et~al.}(2017)Kawashima, Matsumoto, \&
  Matsumoto}]{kawashima2017possible}
Kawashima, T., Matsumoto, Y., \& Matsumoto, R. 2017, Publications of the
  Astronomical Society of Japan, 69, 43

\bibitem[{Kennea {et~al.}(2013)Kennea, Burrows, Kouveliotou, Palmer,
  G{\"o}{\u{g}}{\"u}{\c{s}}, Kaneko, Evans, Degenaar, Reynolds, Miller,
  {et~al.}}]{kennea2013swift}
Kennea, J., Burrows, D., Kouveliotou, C., {et~al.} 2013, The Astrophysical
  Journal Letters, 770, L24

\bibitem[{Kosti{\'c} {et~al.}(2009)Kosti{\'c}, {\v{C}}ade{\v{z}}, Calvani, \&
  Gomboc}]{kostic2009tidal}
Kosti{\'c}, U., {\v{C}}ade{\v{z}}, A., Calvani, M., \& Gomboc, A. 2009,
  Astronomy \& Astrophysics, 496, 307

\bibitem[{Liu \& Melia(2002)}]{liu2002accretion}
Liu, S., \& Melia, F. 2002, The Astrophysical Journal Letters, 566, L77

\bibitem[{Liu {et~al.}(2004)Liu, Petrosian, \& Melia}]{liu2004electron}
Liu, S., Petrosian, V., \& Melia, F. 2004, The Astrophysical Journal Letters,
  611, L101

\bibitem[{Madigan {et~al.}(2016)Madigan, McCourt, \&
  O'Leary}]{madigan2016using}
Madigan, A.-M., McCourt, M., \& O'Leary, R.~M. 2016, Monthly Notices of the
  Royal Astronomical Society, stw2815

\bibitem[{Markoff {et~al.}(2001)Markoff, Falcke, Yuan, \&
  Biermann}]{markoff2001nature}
Markoff, S., Falcke, H., Yuan, F., \& Biermann, P.~L. 2001, Astronomy \&
  Astrophysics, 379, L13

\bibitem[{Marrone {et~al.}(2008)Marrone, Baganoff, Morris, Moran, Ghez,
  Hornstein, Dowell, Munoz, Bautz, Ricker, {et~al.}}]{marrone2008x}
Marrone, D.~P., Baganoff, F., Morris, M., {et~al.} 2008, The Astrophysical
  Journal, 682, 373

\bibitem[{Mori {et~al.}(2013)Mori, Gotthelf, Zhang, An, Baganoff, Barriere,
  Beloborodov, Boggs, Christensen, Craig, {et~al.}}]{mori2013nustar}
Mori, K., Gotthelf, E.~V., Zhang, S., {et~al.} 2013, The Astrophysical Journal
  Letters, 770, L23

\bibitem[{Mossoux \& Grosso(2017)}]{mossoux2017sixteen}
Mossoux, E., \& Grosso, N. 2017, Astronomy \& Astrophysics, 604, A85

\bibitem[{Mossoux {et~al.}(2015)Mossoux, Grosso, Vincent, \&
  Porquet}]{mossoux2015study}
Mossoux, E., Grosso, N., Vincent, F.~H., \& Porquet, D. 2015, Astronomy \&
  Astrophysics, 573, A46

\bibitem[{Mossoux {et~al.}(2016)Mossoux, Grosso, Bushouse, Eckart, Yusef-Zadeh,
  Plambeck, Peissker, Valencia-S, Porquet, Cotton,
  {et~al.}}]{mossoux2016multiwavelength}
Mossoux, E., Grosso, N., Bushouse, H., {et~al.} 2016, Astronomy \&
  Astrophysics, 589, A116

\bibitem[{Neilsen {et~al.}(2013)Neilsen, Nowak, Gammie, Dexter, Markoff,
  Haggard, Nayakshin, Wang, Grosso, Porquet, {et~al.}}]{neilsen2013chandra}
Neilsen, J., Nowak, M., Gammie, C., {et~al.} 2013, The Astrophysical Journal,
  774, 42

\bibitem[{Neilsen {et~al.}(2015)Neilsen, Markoff, Nowak, Dexter, Witzel,
  Barri{\`e}re, Li, Baganoff, Degenaar, Fragile, {et~al.}}]{neilsen2015x}
Neilsen, J., Markoff, S., Nowak, M., {et~al.} 2015, The Astrophysical Journal,
  799, 199

\bibitem[{Nowak {et~al.}(2012)Nowak, Neilsen, Markoff, Baganoff, Porquet,
  Grosso, Levin, Houck, Eckart, Falcke, {et~al.}}]{nowak2012chandra}
Nowak, M., Neilsen, J., Markoff, S., {et~al.} 2012, The Astrophysical Journal,
  759, 95

\bibitem[{Pfuhl {et~al.}(2015)Pfuhl, Gillessen, Eisenhauer, Genzel, Plewa, Ott,
  Ballone, Schartmann, Burkert, Fritz, {et~al.}}]{pfuhl2015galactic}
Pfuhl, O., Gillessen, S., Eisenhauer, F., {et~al.} 2015, The Astrophysical
  Journal, 798, 111

\bibitem[{Ponti {et~al.}(2015)Ponti, De~Marco, Morris, Merloni, Munoz-Darias,
  Clavel, Haggard, Zhang, Nandra, Gillessen, {et~al.}}]{ponti2015fifteen}
Ponti, G., De~Marco, B., Morris, M., {et~al.} 2015, Monthly Notices of the
  Royal Astronomical Society, 454, 1525

\bibitem[{Ponti {et~al.}(2016)Ponti, Jin, De~Marco, Rea, Rau, Haberl,
  Coti~Zelati, Bozzo, Ferrigno, Bower, {et~al.}}]{ponti2016swift}
Ponti, G., Jin, C., De~Marco, B., {et~al.} 2016, Monthly Notices of the Royal
  Astronomical Society, 461, 2688

\bibitem[{Ponti {et~al.}(2017)Ponti, George, Scaringi, Zhang, Jin, Dexter,
  Terrier, Clavel, Degenaar, Eisenhauer, {et~al.}}]{ponti2017powerful}
Ponti, G., George, E., Scaringi, S., {et~al.} 2017, Monthly Notices of the
  Royal Astronomical Society, 468, 2447

\bibitem[{Porquet {et~al.}(2003)Porquet, Predehl, Aschenbach, Grosso, Goldwurm,
  Goldoni, Warwick, \& Decourchelle}]{porquet2003xmm}
Porquet, D., Predehl, P., Aschenbach, B., {et~al.} 2003, Astronomy \&
  Astrophysics, 407, L17

\bibitem[{Porquet {et~al.}(2008)Porquet, Grosso, Predehl, Hasinger,
  Yusef-Zadeh, Aschenbach, Trap, Melia, Warwick, Goldwurm,
  {et~al.}}]{porquet2008x}
Porquet, D., Grosso, N., Predehl, P., {et~al.} 2008, Astronomy \& Astrophysics,
  488, 549

\bibitem[{Quataert(2002)}]{quataert2002thermal}
Quataert, E. 2002, The Astrophysical Journal, 575, 855

\bibitem[{Quataert(2003)}]{quataert2003radiatively}
---. 2003, Astronomische Nachrichten: Astronomical Notes, 324, 435

\bibitem[{Reid \& Brunthaler(2004)}]{reid2004proper}
Reid, M.~J., \& Brunthaler, A. 2004, The Astrophysical Journal, 616, 872

\bibitem[{Scargle {et~al.}(2013)Scargle, Norris, Jackson, \&
  Chiang}]{scargle2013studies}
Scargle, J.~D., Norris, J.~P., Jackson, B., \& Chiang, J. 2013, The
  Astrophysical Journal, 764, 167

\bibitem[{Schartmann {et~al.}(2012)Schartmann, Burkert, Alig, Gillessen,
  Genzel, Eisenhauer, \& Fritz}]{schartmann2012simulations}
Schartmann, M., Burkert, A., Alig, C., {et~al.} 2012, The Astrophysical
  Journal, 755, 155

\bibitem[{Shakura \& Sunyaev(1973)}]{shakura1973black}
Shakura, N.~I., \& Sunyaev, R.~A. 1973, Astronomy and Astrophysics, 24, 337

\bibitem[{Verner {et~al.}(1996)Verner, Ferland, Korista, \&
  Yakovlev}]{verner1996atomic}
Verner, D., Ferland, G.~J., Korista, K., \& Yakovlev, D. 1996, arXiv preprint
  astro-ph/9601009

\bibitem[{Wang {et~al.}(2013)Wang, Nowak, Markoff, Baganoff, Nayakshin, Yuan,
  Cuadra, Davis, Dexter, Fabian, {et~al.}}]{wang2013dissecting}
Wang, Q., Nowak, M., Markoff, S., {et~al.} 2013, Science, 341, 981

\bibitem[{Williams {et~al.}(2017)Williams, Clavel, Newton, \&
  Ryzhkov}]{williams2017pwkit}
Williams, P.~K., Clavel, M., Newton, E., \& Ryzhkov, D. 2017, Astrophysics
  Source Code Library

\bibitem[{Wilms {et~al.}(2000)Wilms, Allen, \& McCray}]{wilms2000absorption}
Wilms, J., Allen, A., \& McCray, R. 2000, The Astrophysical Journal, 542, 914

\bibitem[{Witzel {et~al.}(2012)Witzel, Eckart, Bremer, Zamaninasab,
  Shahzamanian, Valencia-S, Sch{\"o}del, Karas, Lenzen, Marchili,
  {et~al.}}]{witzel2012source}
Witzel, G., Eckart, A., Bremer, M., {et~al.} 2012, The Astrophysical Journal
  Supplement Series, 203, 18

\bibitem[{Xu {et~al.}(2006)Xu, Narayan, Quataert, Yuan, \&
  Baganoff}]{xu2006thermal}
Xu, Y.-D., Narayan, R., Quataert, E., Yuan, F., \& Baganoff, F.~K. 2006, The
  Astrophysical Journal, 640, 319

\bibitem[{Yuan \& Narayan(2014)}]{yuan2014hot}
Yuan, F., \& Narayan, R. 2014, Annual Review of Astronomy and Astrophysics, 52,
  529

\bibitem[{Yuan {et~al.}(2003)Yuan, Quataert, \& Narayan}]{yuan2003nonthermal}
Yuan, F., Quataert, E., \& Narayan, R. 2003, The Astrophysical Journal, 598,
  301

\bibitem[{Yuan \& Wang(2016)}]{yuan2015systematic}
Yuan, Q., \& Wang, Q.~D. 2016, Monthly Notices of the Royal Astronomical
  Society, 456, 1438

\bibitem[{Yuan {et~al.}(2018)Yuan, Wang, Liu, \& Wu}]{yuan2017systematic}
Yuan, Q., Wang, Q.~D., Liu, S., \& Wu, K. 2018, Monthly Notices of the Royal
  Astronomical Society, 473, 306

\bibitem[{Zhang {et~al.}(2017)Zhang, Baganoff, Ponti, Neilsen, Tomsick, Dexter,
  Clavel, Markoff, Hailey, Mori, {et~al.}}]{zhang2017sagittarius}
Zhang, S., Baganoff, F.~K., Ponti, G., {et~al.} 2017, The Astrophysical
  Journal, 843, 96

\bibitem[{Zubovas {et~al.}(2012)Zubovas, Nayakshin, \&
  Markoff}]{zubovas2012sgr}
Zubovas, K., Nayakshin, S., \& Markoff, S. 2012, Monthly Notices of the Royal
  Astronomical Society, 421, 1315

\end{thebibliography}

\end{document}